\documentclass[twocolumn,10pt,cleanfoot]{asme2e}
\usepackage[italicdiff]{physics}
\usepackage{graphicx}
\usepackage{subfigure}

\newcommand{\theo}{\theta_{\circ}}
\newcommand{\phio}{\phi_{\circ}}

\confshortname{IDETC/CIE 2021}
\conffullname{the ASME 2021 International Design Engineering Technical Conferences \&\\
              Computers and Information in Engineering Conference}

\confdate{August 17-20, 2021, Virtual, Online}
\confyear{2021}
\confcity{Online}
\confcountry{Virtual}

\papernum{DETC2021/72184}

\title{On a Class of Polar Log-Aesthetic Curves}

\author{Victor Parque
    \affiliation{
	System Design Laboratory\\
	Department of Modern Mechanical Engineering\\
	Waseda University\\
	Tokyo, 169-8555\\
    Email: parque@aoni.waseda.jp
    }	
}

\begin{document}

\maketitle

\begin{abstract}

Curves are essential concepts that enable compounded aesthetic curves, e.g., to assemble complex silhouettes, match a specific curvature profile in industrial design, and construct smooth, comfortable, and safe trajectories in vehicle-robot navigation systems. New mechanisms able to encode, generate, evaluate, and deform aesthetic curves are expected to improve the throughput and the quality of industrial design. In recent years, the study of (log) aesthetic curves have attracted the community's attention due to its ubiquity in natural phenomena such as bird eggs, butterfly wings, falcon flights, and manufactured products such as Japanese swords and automobiles.

A (log) aesthetic curve renders a logarithmic curvature graph approximated by a straight line, and polar aesthetic curves enable to mode user-defined dynamics of the polar tangential angle in the polar coordinate system. As such, the curvature profile often becomes a by-product of the tangential angle.

In this paper, we extend the concept of polar aesthetic curves and establish the analytical formulations to construct aesthetic curves with user-defined criteria. In particular, we propose the closed-form analytic characterizations of polar log-aesthetic curves meeting user-defined criteria of curvature profiles and dynamics of polar tangential angles. We present numerical examples portraying the feasibility of rendering the logarithmic curvature graphs represented by a straight line. Our approach enables the seamless characterization of aesthetic curves in the polar coordinate system, which can model aesthetic shapes with desirable aesthetic curvature profiles.

\end{abstract}



\section{INTRODUCTION}

Curves are ubiquitous entities in industrial design, enabling the creation of functional and aesthetic machines and consumer electronics\cite{sequin05,miura}, and the smooth and energy-efficient motion planning in robots\cite{yong04,can13}. The research on conceptual mechanisms allowing to construct aesthetic curves is relevant, not only to render sketches to enable designers to extrapolate ideas to accurate symbolic expressions in computers but also to facilitate the characterization of the first principles of aesthetics in geometric design\cite{sequin09,miura14}. Thus, studying mechanisms to generate, evaluate, and deform aesthetic curves is expected to improve the throughput and the quality of the industrial design process.

In this paper we focus our attention on a class of curves defined by spirals, which are relevant for applications in efficient coverage path planning\cite{hassan18,lee11,yong04,hassan20}, folding membranes into compact structures\cite{pelle92,pelle01,nojima01a,nojima01b,parque20}, and to devise new interpolation patterns for metaheuristics\cite{guo20,mah20}, among many other applications. For instance, in \cite{hassan18}, an archimedean spiral curve is used to generate spiral paths within a circle, and then the generated paths are analytically mapped to a bounding rectangle; and in \cite{hassan20}, the archimedean spiral is deformed into a squircircle (a shape intermediary between a circle and a square) for efficient coverage planning. Furthermore, in \cite{lee11}, an archimedean-like formulation is used to sample path points for coverage, whose configuration is filtered and optimized. Generally speaking spiral-based paths (robotics) and spiral-based folding configurations (membranes applications) are desirable due to the fact of avoiding frequent or aggressive changes in acceleration, allowing the smooth deployment of mobile robots\cite{hassan20,parquesmooth20} and membrane structures\cite{pelle92,pelle01,nojima01a,nojima01b,parque20}.

The study of spirals has received considerable attention ubiquitously. On the book "On Growth and Form", D. W. Thompson discussed the natural tendency of growth in terms of logarithmic spirals in conch, horn, tusk, and others\cite{thomson}. Also, in the book "The Curves of Life", T. A. Cook studied various spiral curves in nature, science and art\cite{curvesoflife}. It is relevant to note that the equiangular property of the logarithmic spiral is found in the shell of the chambered nautilus and the flight patterns of peregrine falcons, hawks and eagles\cite{tuck00}. In particular, hawks and falcons resolve the conflict between limited field of vision and challenging aerodynamics by holding their head straight and by flying along a path with a logarithmic spiral pattern, as such they not only move quickly and smoothly but also they keep the line of sight at maximal acuity pointed sideways at the prey\cite{tuck00,tuck00b}.

In line with the above, Harada studied the class of curves appearing in natural entities such as bird's eggs, butterflies' wings, and man-made products such as Japanese swords and automobile shapes, and coined the term of \emph{aesthetic curves} to define the class of curves whose logarithmic distribution of curvature is approximated by a straight line\cite{harada}. Miura proposed the analytical equations of aesthetic curves by using generalizations of the logarithmic spiral and the clothoid spiral, and studied the properties of the \emph{logarithmic curvature graph}\cite{miura,yoshida10}. Here, the horizontal axis measures $\log \rho$ and the vertical axis measures $\log | dL/ d \log \rho |$, where $L$ is arc length of the curve and $\rho$ is its radius of curvature.

The fundamental principle behind an aesthetic curve is that it renders a logarithmic curvature graph approximated by a straight line, and due to the fact of using the logarithmic expression of curvature, the aesthetic curves was later coined as \emph{log-aesthetic curves}. The basic intuition behind rendering the aesthetic curve is that the eye is more sensitive to relative variations of curvature than absolute ones; thus, large variation of curvature is tolerable in regions of high curvature. In contrast, even minor variations of curvature become visible in low curvature regions\cite{sequin09}. Examples of log-aesthetic curves involve the logarithmic spiral, the clothoid, the involute curves, among others\cite{miu12,su18,su218}. More recently, they have been characterized by linear regression on curves and surfaces\cite{higashi14a,higashi14b,higashi15,parque16}, by incomplete Gamma functions\cite{yoshida12}, and by stationary integrable flows on plane curves governed by the Burgers equation\cite{ino18}.

\begin{figure}[t]
\begin{center}
\subfigure[Curve]{\includegraphics[width=0.48\columnwidth]{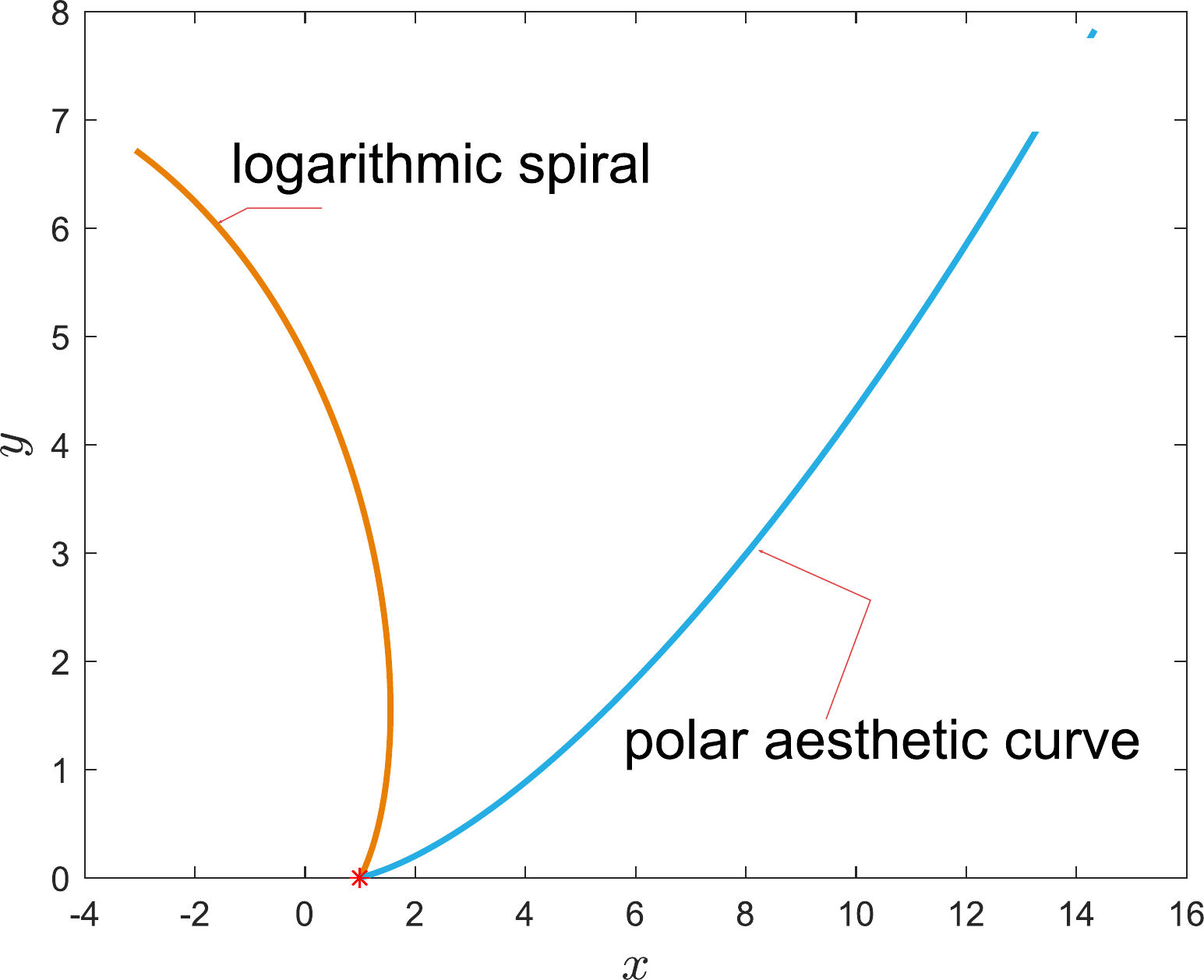}}
\subfigure[Logarithmic curvature]{\includegraphics[width=0.48\columnwidth]{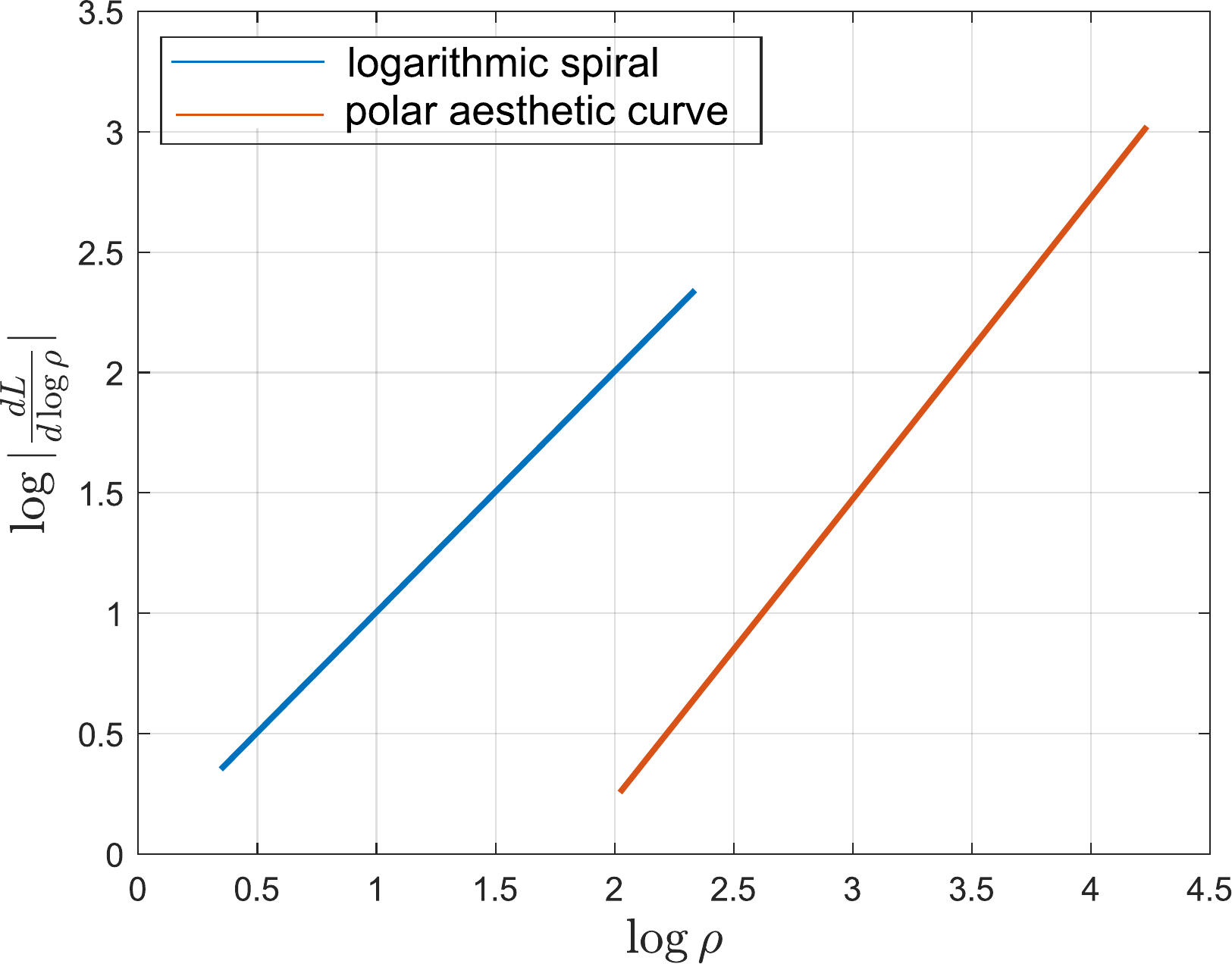}}
\end{center}
\caption{Comparison of the logarithmic spiral with $\phi = \pi/4$ and the polar aesthetic curve with $\phi = 0.2\theta + \pi/24$.}
\label{sample}
\end{figure}

By extending the logarithmic spiral for variable polar tangential angles, Miura proposed the concept of the \emph{polar aesthetic curve} to define the class of aesthetic curves in which the angle between the tangent and the radial direction is either a linear function or an arbitrary function of the radial angle\cite{miura13,miura15}. Fig. \ref{sample} shows an example of a polar aesthetic curve when the polar tangential angle $\phi$ is a linear function of the polar angle $\theta$. Since the polar aesthetic curve is derived from the logarithmic spiral, its logarithmic curvature graph is represented by a line, as shown by Fig. \ref{sample}-(b). Thus, by choosing the dynamics of the tangential angle, it becomes possible to design aesthetic curves, in which the curvature profile is a byproduct of the tangential angle $\phi$.

In this paper, having in mind the potential applications to design compounded aesthetic curves, to assemble composition of complex silhouettes, and to design curves that match a specific curvature profile in the polar coordinates, we extend the polar notion of polar aesthetic curves and establish the analytical formulations to construct curves with user-defined curvature profiles and polar tangential angle dynamics. In particular, we derive the closed-form properties that describe the relations between radius, polar angle, arc length, and radius of curvature derived from user-defined criteria in both the curvature profile and the polar tangential angle. Our proposed approach enables us to construct polar aesthetic curves whose logarithmic distribution of curvature is expressed by a controllable straight line, thus maintaining the fundamental criterion of aesthetic curves. As such, our approach enables the seamless characterization of log-aesthetic curves whose curvature profiles are completely defined by the designer. To the best of our knowledge, our proposed formulations are the first presented in the field.

\section{PRELIMINARIES}

Here, we describe a number of concepts being relevant to our approach.

\subsection{Polar Curves}

\begin{figure}[t]
\begin{center}
{\includegraphics[width=0.98\columnwidth]{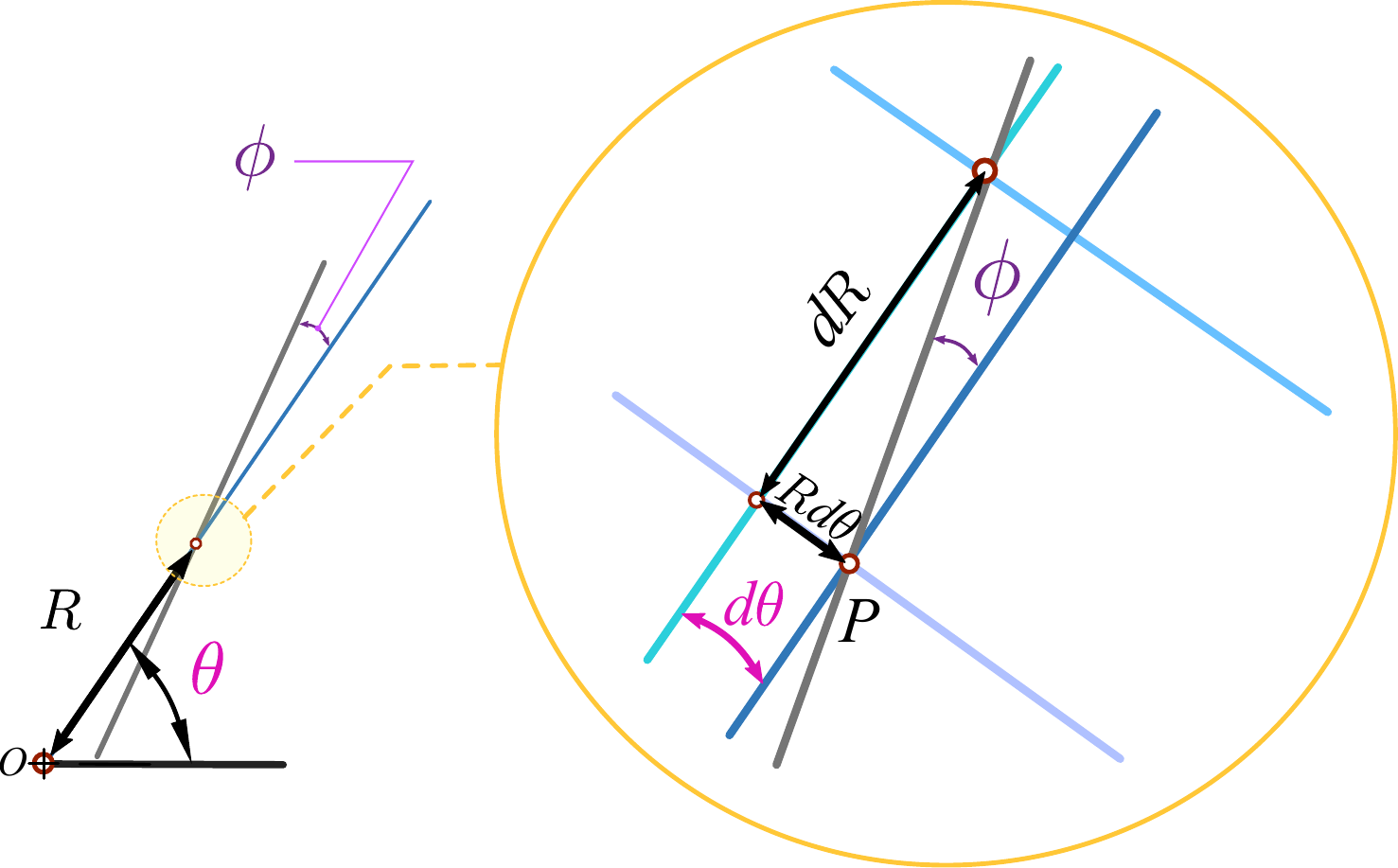}}
\end{center}
\caption{Basic polar coordinate system.}
\label{cor}
\end{figure}

When a curve is defined in polar coordinates as shown by Fig. \ref{cor}, we can define the following relation from the infinitesimal region

\begin{equation}\label{tanphi}
  \tan \phi = \frac{R \dd \theta}{\dd R},
\end{equation}
where $\theta$ is the polar angle (from the $x$ axis), $R$ is the radial distance, and $\phi$ is the angle between the tangent and the radial direction. Also, when $R$ is a function of $\theta$, the curvature $\kappa$ is given by

\begin{equation}\label{kappa}
  \kappa = \frac{ R^2 + 2R'^2 - R.R'' }{(R^2 + R'^2)^{\frac{3}{2}}},
\end{equation}
where $R'$ denotes the derivative of $R$ with respect to $\theta$. We can combine Eqn. (\ref{tanphi}) and Eqn. (\ref{kappa}) and obtain


\begin{equation}\label{sinr}
  \frac{\sin \phi \Big ( 1 + \displaystyle \dv{\phi}{\theta} \Big )}{R} = \frac{1}{\rho},
\end{equation}
where $\rho$ is the radius of curvature. Also, by definition of the infinitesimal region as shown by Fig. \ref{cor}, we can obtain the following relationships

\begin{equation}\label{sinphi}
\sin \phi = \frac{R \dd \theta}{\dd L},
\end{equation}

\begin{equation}\label{curv}
  \Big ( 1+ \dv{\phi}{\theta} \Big ) \dv{\theta}{L} = \frac{1}{\rho}
\end{equation}

\begin{equation}\label{dbeta}
  \displaystyle \dv{\theta}{L} + \dv{\phi}{L} = \dv{\beta}{L} = \frac{1}{\rho},
\end{equation}
where $L$ denotes the arc length, $\beta$ denotes the tangential angle. By combining Eqn. (\ref{sinr}) and Eqn. (\ref{dbeta}), we obtain the following

\begin{equation}\label{RL}
  R = \displaystyle \frac{\sin \phi}{\displaystyle \frac{1}{\rho} - \dv{\phi}{L}}
\end{equation}

\begin{figure}[t]
\begin{center}
{\includegraphics[width=0.98\columnwidth]{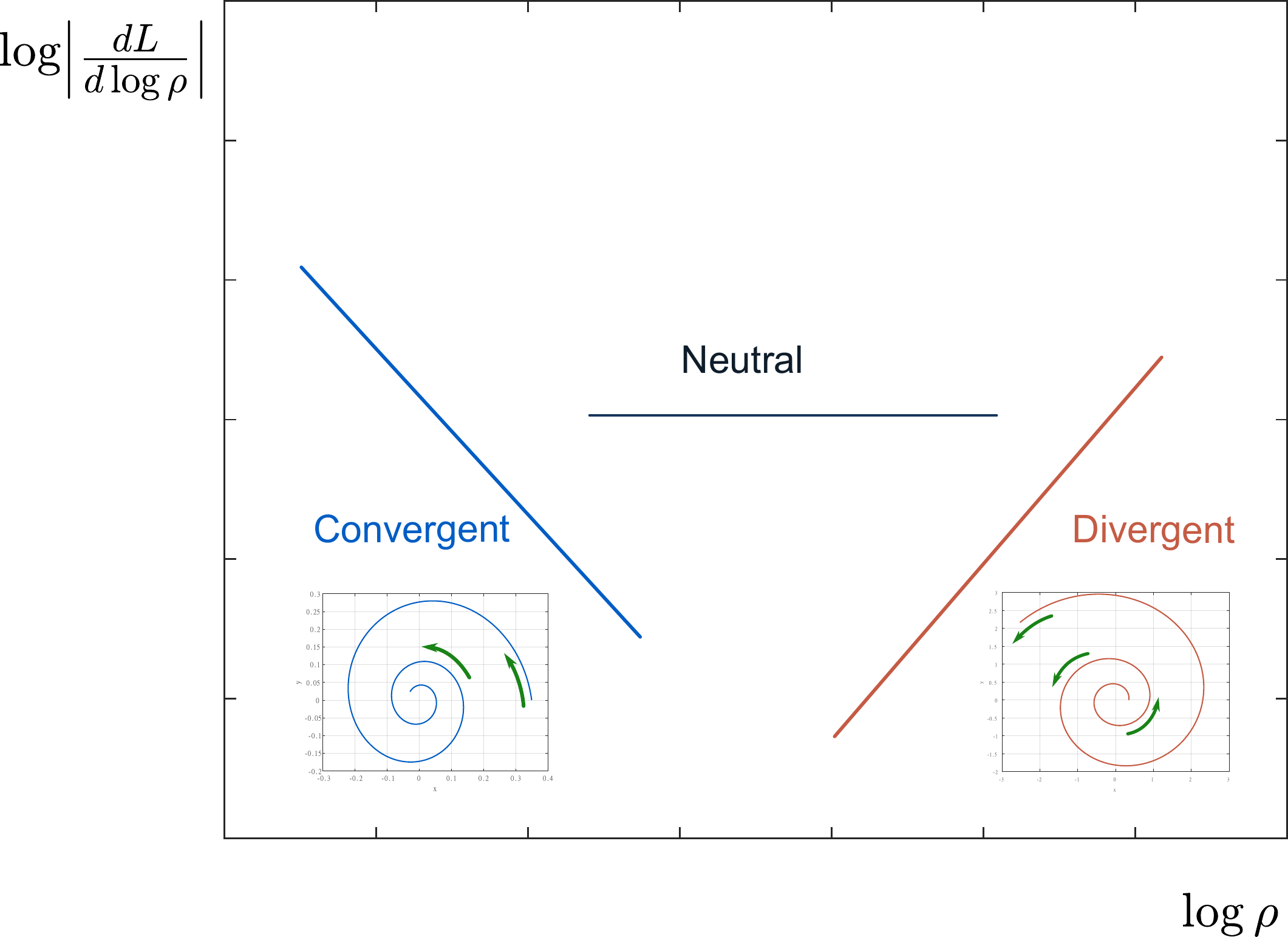}}
\end{center}
\caption{Basic concept of a logarithmic curvature graph.}
\label{logcurva}
\end{figure}

\begin{figure*}[h]
\begin{center}
\subfigure[Curve]{\includegraphics[width=0.32\textwidth]{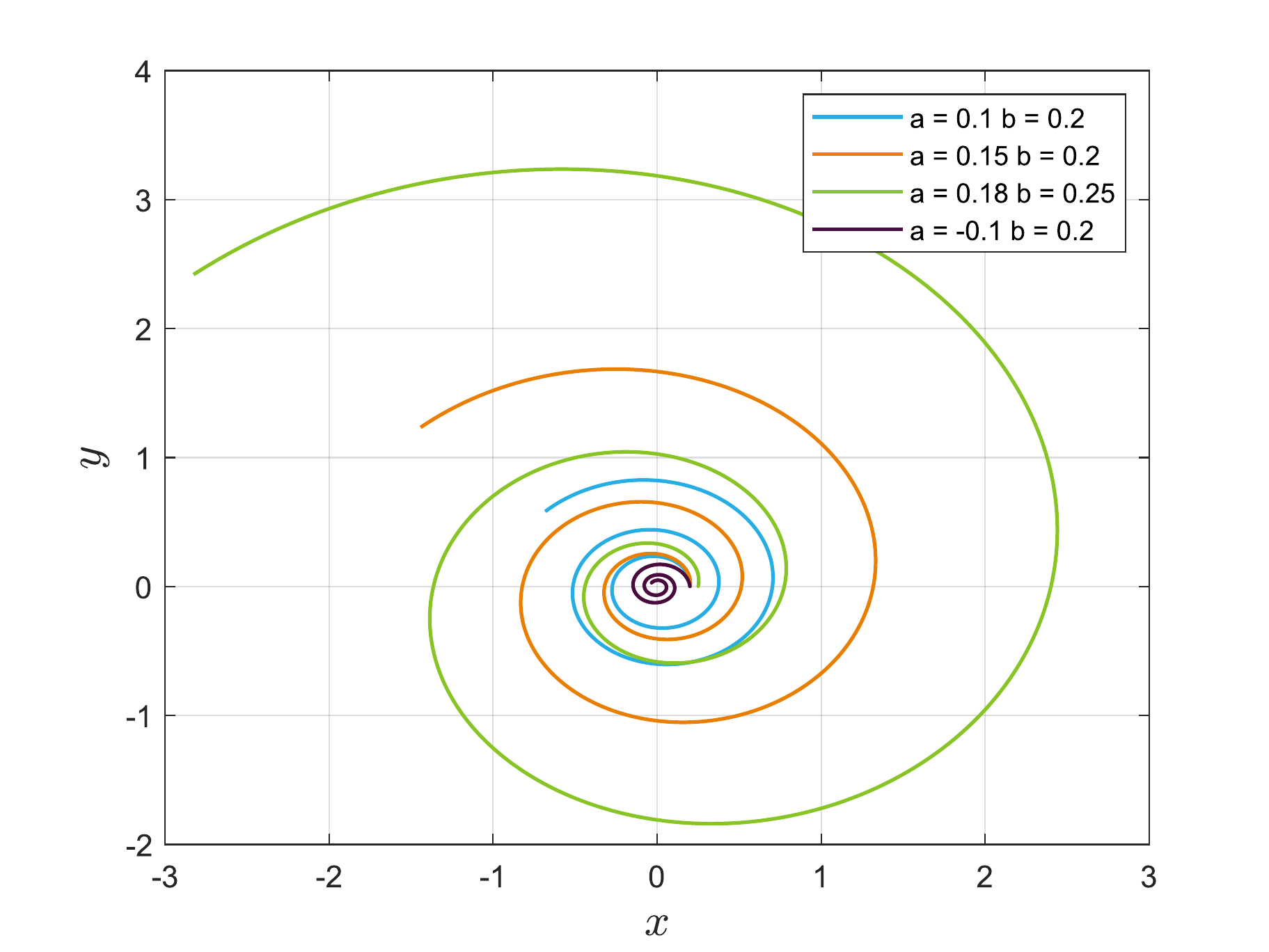}}
\subfigure[Radius of curvature]{\includegraphics[width=0.32\textwidth]{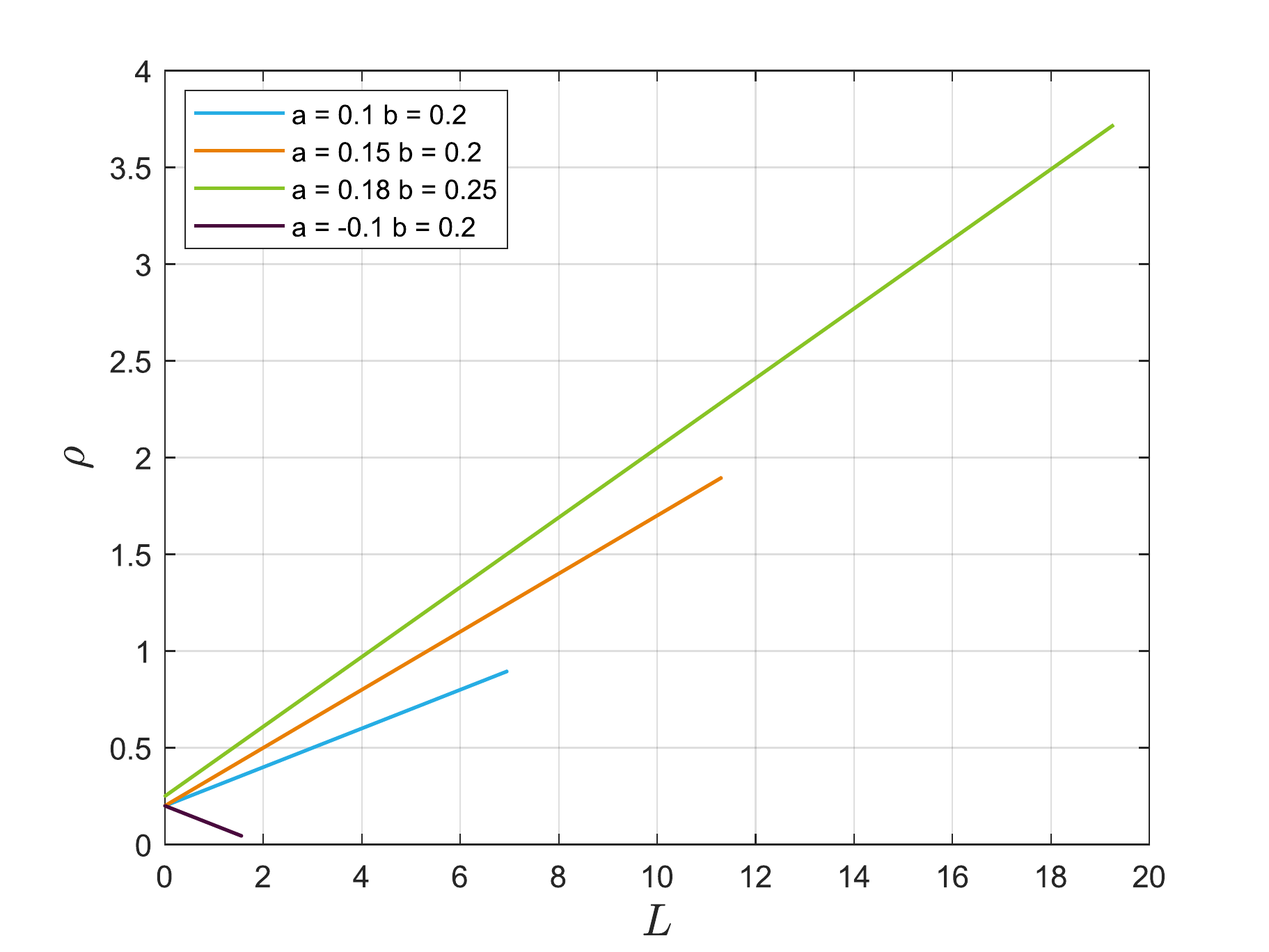}}
\subfigure[Logarithmic curvature]{\includegraphics[width=0.32\textwidth]{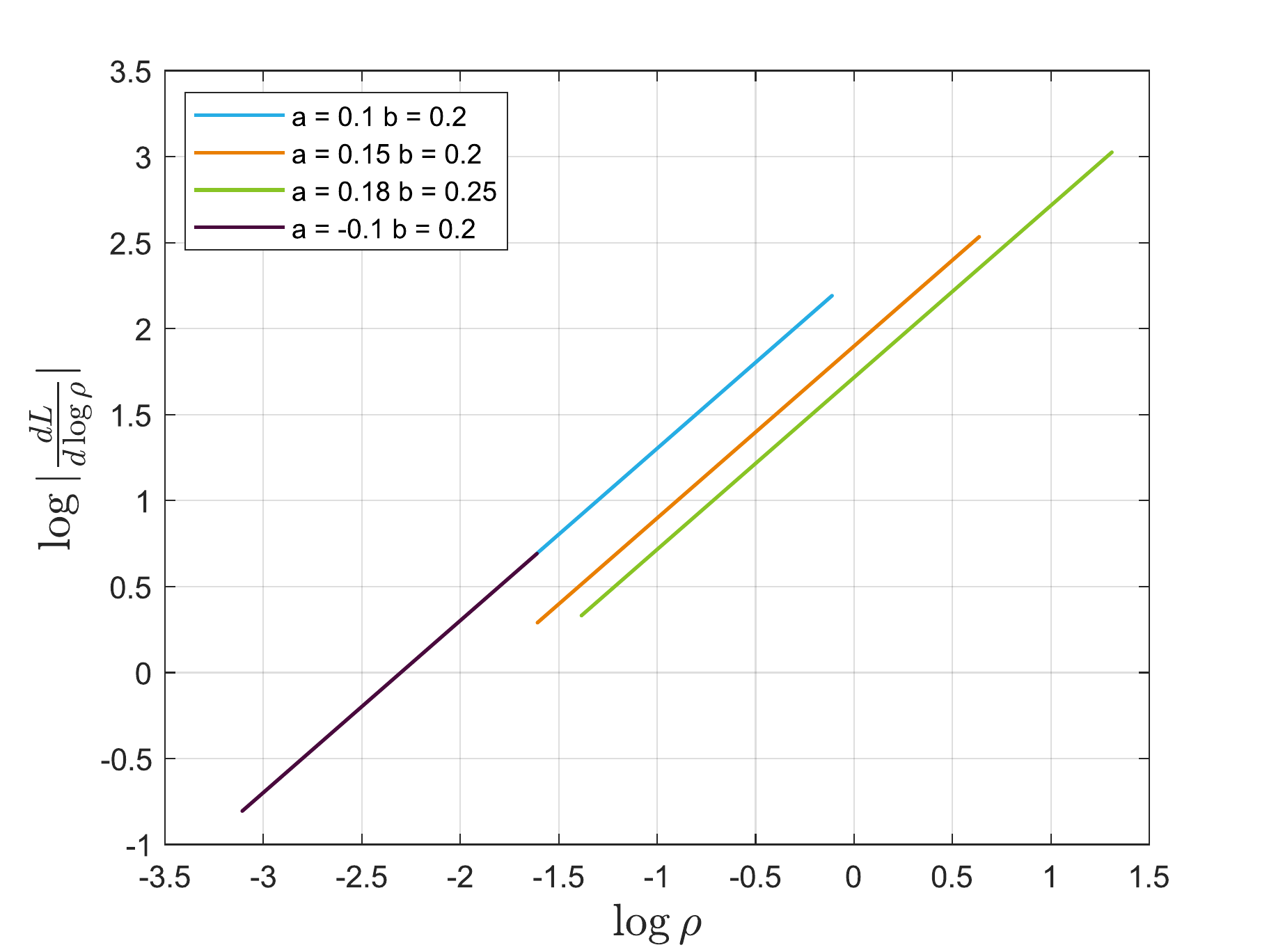}}
\end{center}
\caption{Log aesthetic curves with $n = 1$, $\phi = \pi/2$, $\theta \in [0, 15]$ rad.}
\label{curveA1}
\end{figure*}

\subsection{Log-Aesthetic Curves}

The log-aesthetic curves satisfy the following equation\cite{miura}:

\begin{equation}\label{roun}
  \rho^n = aL + b,
\end{equation}
where $n \neq 0,~ a, ~ b$ are real constants. When $n= 1$ and $n= -1$, it is possible to model two typical aesthetic curves: the logarithmic spiral and the clothoid curves, respectively. The above Eqn. (\ref{roun}) is inspired by the curves whose logarithmic distribution of curvature is approximated by a straight line. Harada\cite{harada} coined the term of \emph{aesthetic curves} to denote the class of such curves appearing in Japanese swords, bird's eggs and butterflies' wings; and Miura proposed the analytical equations of log-aesthetic curves using a \emph{logarithmic curvature graph}\cite{miura,yoshida10}, in which the horizontal axis measures $\log \rho$ and the vertical axis measures $\log | dL/ d \log \rho |$. Basically, the fundamental equation of the log-aesthetic curves portrays a logarithmic curvature graph with a straight line whose slope is equal to $n$, as follows

\begin{equation}\label{logaes}
  \log \Big | \dv{L}{\log \rho} \Big | = n\log \rho + C,
\end{equation}
where $C$ is a constant, and $n$ becomes the slope of the logarithmic curvature graph. As such, a positive (negative) slope of the logarithmic curvature graph indicates the convergence (divergence) towards (from) the origin and provides either a centripetal (sharp) or decelerating (accelerating) impression. For instance, in a study of more than one hundred sample curves drawn by expert designers, Harada et al.\cite{harada99} observed that Italian car designers preferred bonnets consisting of accelerating curves such as the ones with slope $n = -1$, whereas Japanese car designers preferred the decelerating curves such as the parabola.

Although Eqn. (\ref{logaes}) is expressed as a planar curve parameterized by arc length, it is possible to extrapolate the above-mentioned concept to space curves encoded by any differential parametric equation such as B\'{e}zier and NURBS, and thus establish criteria of continuity of compound curves\cite{yoshida10}.

\section{POLAR AESTHETIC CURVES}

Based on the discussions mentioned above, in this section we describe the polar log-aesthetic curves. The notion of \emph{polar aesthetic curves} was coined by Miura\cite{miura13,miura15} to characterize the class of curves in which the angle between the tangent and the radial direction is related to the radial angle for a scissor design application, for which criteria of monotonicity was established. In the following, we extend the above concept and establish the analytical criteria to construct polar log-aesthetic curves while ensuring the fundamental criteria of linear logarithmic curvature. Due to the nature of Eqn. (\ref{roun}), we divide our analysis into two groups, as follows.

\subsection{Class I. } When $n = 1$, we combine Eqn. (\ref{dbeta}) and Eqn. (\ref{roun}) and obtain
\begin{equation}\label{A}
  \dd \theta + \dd \phi = \displaystyle \frac{\dd L}{aL+b}.
\end{equation}

By integrating (\ref{A}) over the respective domains $[\theo, \theta]$, $[\phio, \phi]$ and length $[0, L]$, we obtain

\begin{equation}\label{A1}
  \theta - \theo + \phi - \phio = \frac{1}{a}\ln \Big (  \frac{aL}{b} + 1 \Big ).
\end{equation}

And when $\phi$ is a function of $\theta$, such as $\phi = f(\theta)$, we can compute the arc length $L$ from (\ref{A1}) by

\begin{equation}\label{LA}
  L = \frac{b}{a} \Bigg ( \exp \Big ({a(  \theta - \theo + f(\theta) - f(\theo) )} \Big )  - 1 \Bigg ),
\end{equation}
and by using (\ref{sinr}) the radius $R$ becomes
\begin{equation}\label{RA}
  R = \big (aL +b\big) \big (1 + f'(\theta)\big  ) \sin f(\theta),
\end{equation}
where $f'$ is the derivative with respect to $\theta$, and $\rho = aL +b$. In the following, we study two particular cases.

\subsubsection{When $\phi$ is constant} Here we assume that $\phi$ is a real constant. Thus,
 Eqn. (\ref{LA}) and Eqn. (\ref{RA}) become

\begin{equation}\label{LA1}
  L = \frac{b}{a} \Big ( \exp \big ({a(  \theta - \theo )} \big )  - 1 \Big ),
\end{equation}
\begin{equation}\label{RA1}
  R = (aL +b) \sin \phi.
\end{equation}

The above denotes the class of logarithmic spiral curves, in which the angle $\phi$ is a constant. An example when $n = 1$, $\phi = \pi/2$, $\theta \in [0, 15]$ rad. is provided in Fig. \ref{curveA1}.

\begin{figure*}[t]
\begin{center}
\subfigure[Curve]{\includegraphics[width=0.32\textwidth]{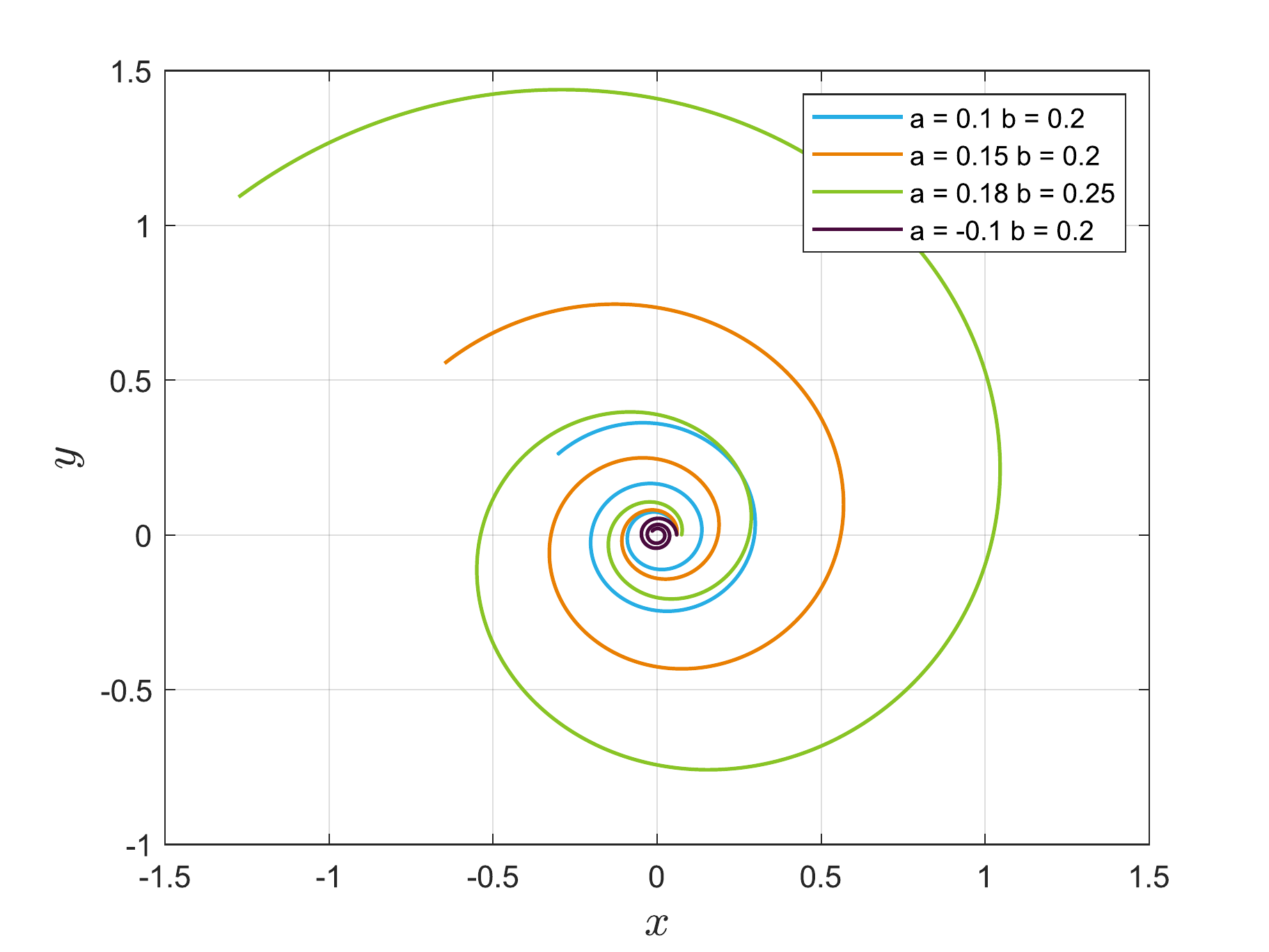}}
\subfigure[Radius of curvature]{\includegraphics[width=0.32\textwidth]{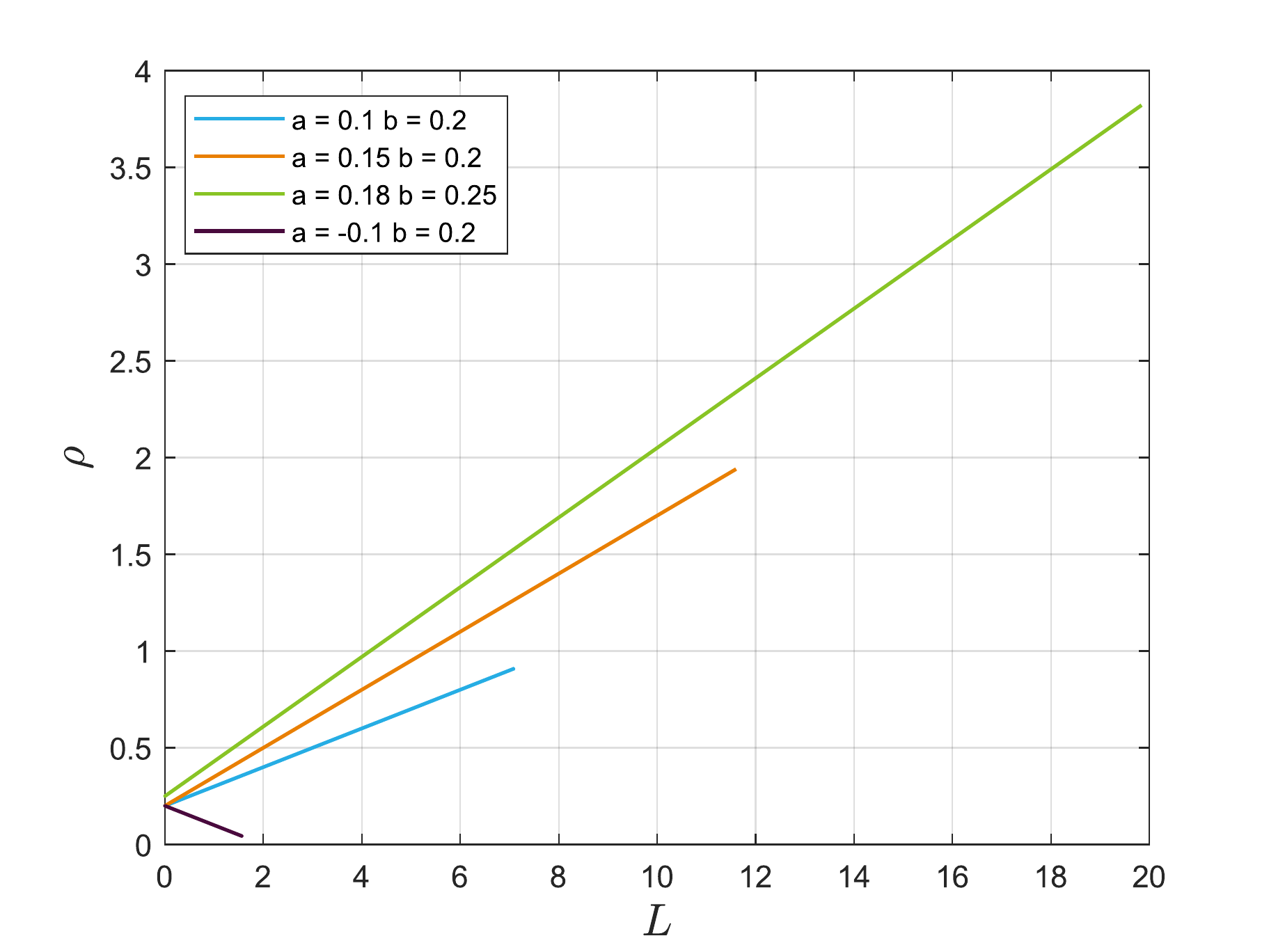}}
\subfigure[Logarithmic curvature]{\includegraphics[width=0.32\textwidth]{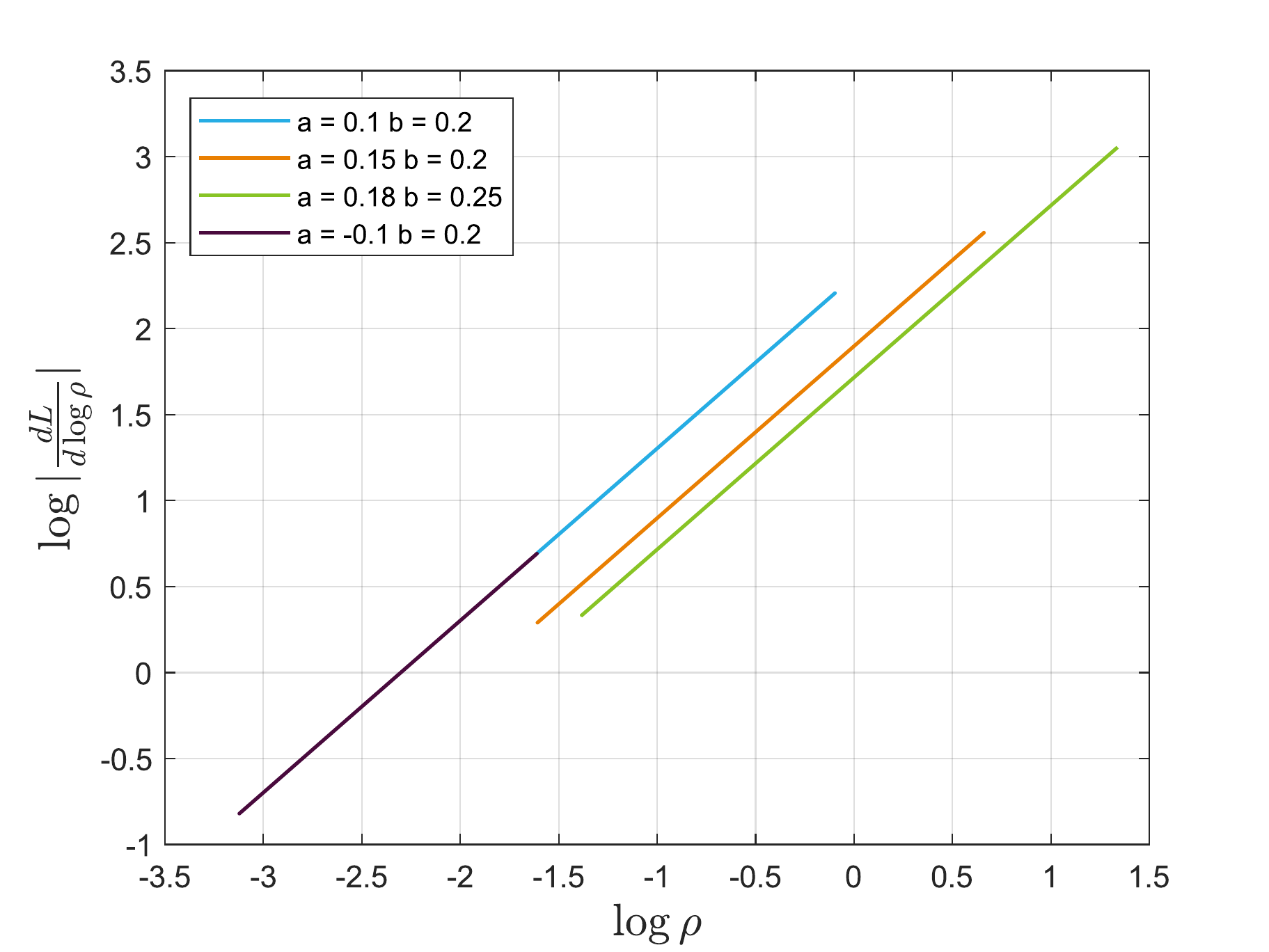}}
\end{center}
\caption{Curves with $n = 1$, $\phi = 0.01\theta + 0.3$, $\theta \in [0, 15]$ rad.}
\label{curveA2}
\end{figure*}

\subsubsection{When $\phi$ is a linear function of $\theta$}
Here we assume that $\phi$ is a encoded by the following function $\phi = c_1\theta + c_0 $. Thus,
 Eqn. (\ref{LA}) and Eqn. (\ref{RA}) become

\begin{equation}\label{LA2}
  L = \frac{b}{a} \Bigg ( \exp \Big ({a(1+c_1)( \theta - \theo )} \Big )  - 1 \Bigg ),
\end{equation}

\begin{equation}\label{RA2}
  R = (aL+b) (1 + c_1 )  \sin (c_1\theta + c_0).
\end{equation}

An example when $n = 1$, $\phi = 0.01\theta + 0.3$, $\theta \in [0, 15]$ rad. is provided in Fig. \ref{curveA1}.

\begin{figure*}[h]
\begin{center}
\subfigure[Curve]{\includegraphics[width=0.32\textwidth]{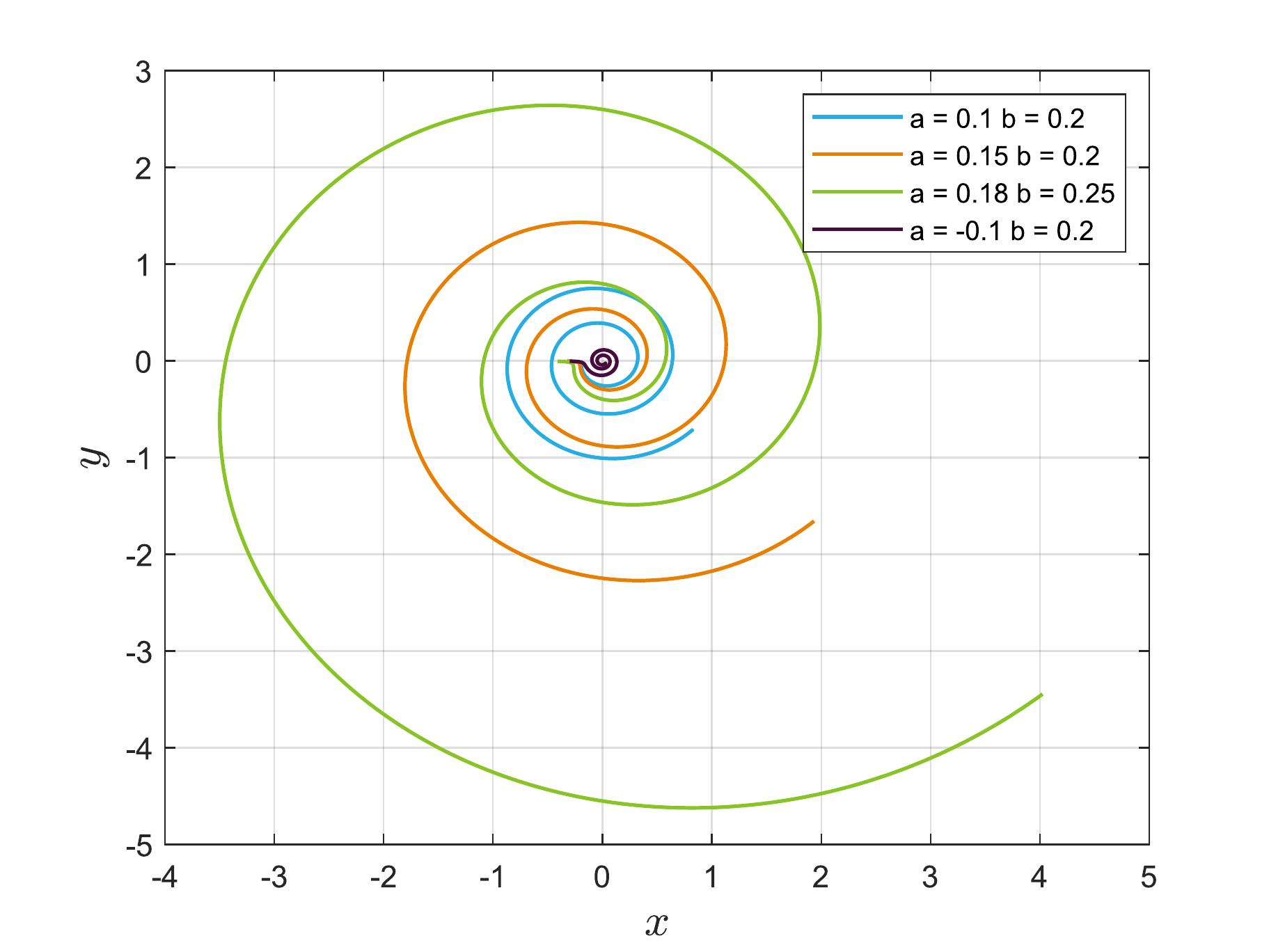}}
\subfigure[Radius of curvature]{\includegraphics[width=0.32\textwidth]{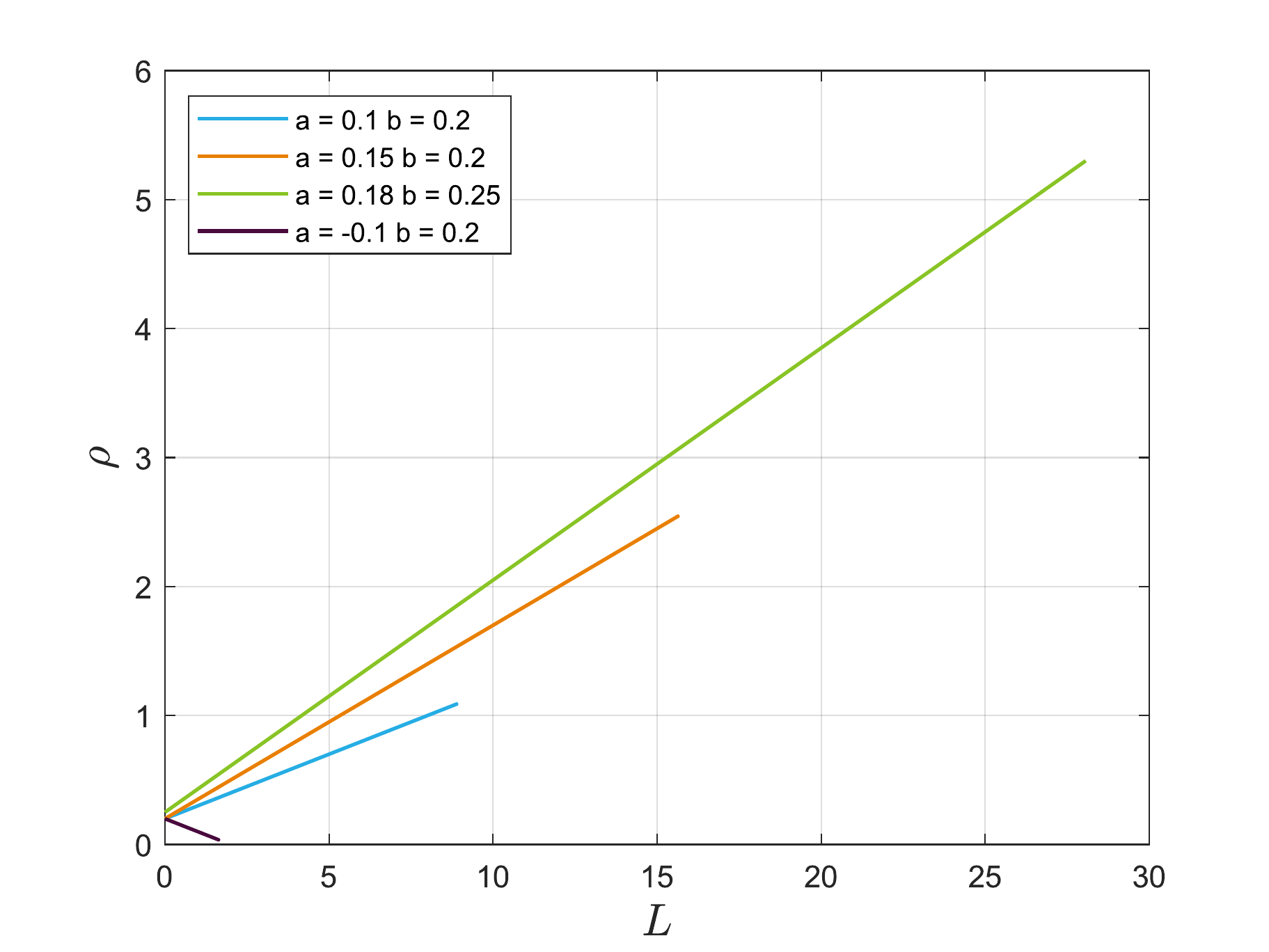}}
\subfigure[Logarithmic curvature]{\includegraphics[width=0.32\textwidth]{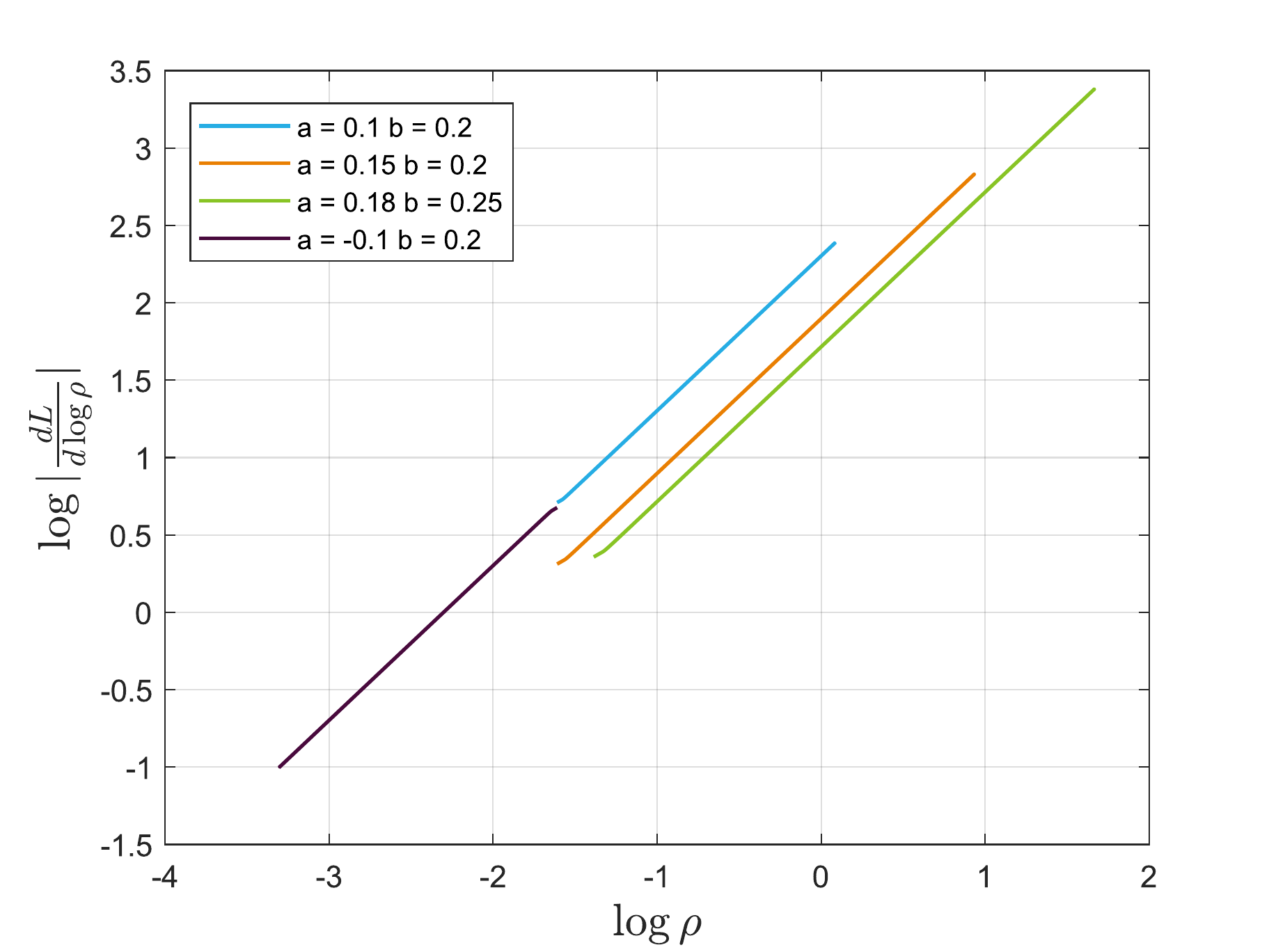}}
\end{center}
\caption{Polar aesthetic curves for arbitrary function of $\phi$ when $n = 1$, $\phi = \theta^\frac{1}{4} + 3$, $\theta \in [0, 15]$ rad.}
\label{curveAg}
\end{figure*}

By using Eq. \ref{LA} and Eqn. \ref{RA}, it is also possible to model polar aesthetic curves considering arbitrary functions of $\phi$. For instance, an example portraying the class of curves when $n = 1$, $\phi = \theta^\frac{1}{4} + 3$, $\theta \in [0, 15]$ rad. is provided in Fig. \ref{curveAg}.

\subsection{Class II}

When $n \neq 1$, we combine Eqn. (\ref{dbeta}) and Eqn. (\ref{roun}) and obtain
\begin{equation}\label{B}
  \dd \theta + \dd \phi = \displaystyle \frac{\dd L}{\displaystyle (aL+b)^{\frac{1}{n}}}.
\end{equation}

By integrating (\ref{B}) over the respective domains $[\theo, \theta]$, $[\phio, \phi]$ and length $[0, L]$, we obtain

\begin{equation}\label{B1}
  \theta - \theo + \phi - \phio = \frac{n}{a(n-1)} \Big ( (aL+b)^{1-\frac{1}{n}} -b^{1-\frac{1}{n}} \Big ).
\end{equation}

When $\phi$ is a function of $\theta$, such as $\phi = f(\theta)$, we can compute the arc length $L$ from (\ref{A1}) by

\begin{equation}\label{LB}
  L = \frac{1}{a} \Bigg ( \Bigg (\frac{a \big (n-1 \big ) \big (\theta - \theo + f(\theta) - f(\theo) \big ) + nb^{ 1-\frac{1}{n} }}{n} \Bigg )^{\frac{n}{n-1}} - b \Bigg ),
\end{equation}
and by using (\ref{sinr}) the radius $R$ becomes
\begin{equation}\label{RB}
  R = \big (aL +b \big )^{\frac{1}{n}} \big (1 + f'(\theta) \big ) \sin f(\theta),
\end{equation}
where $f'$ is the derivative with respect to $\theta$, and $\rho = (aL +b)^{\frac{1}{n}} $.

\begin{figure*}[t]
\begin{center}
\subfigure[Curve]{\includegraphics[width=0.32\textwidth]{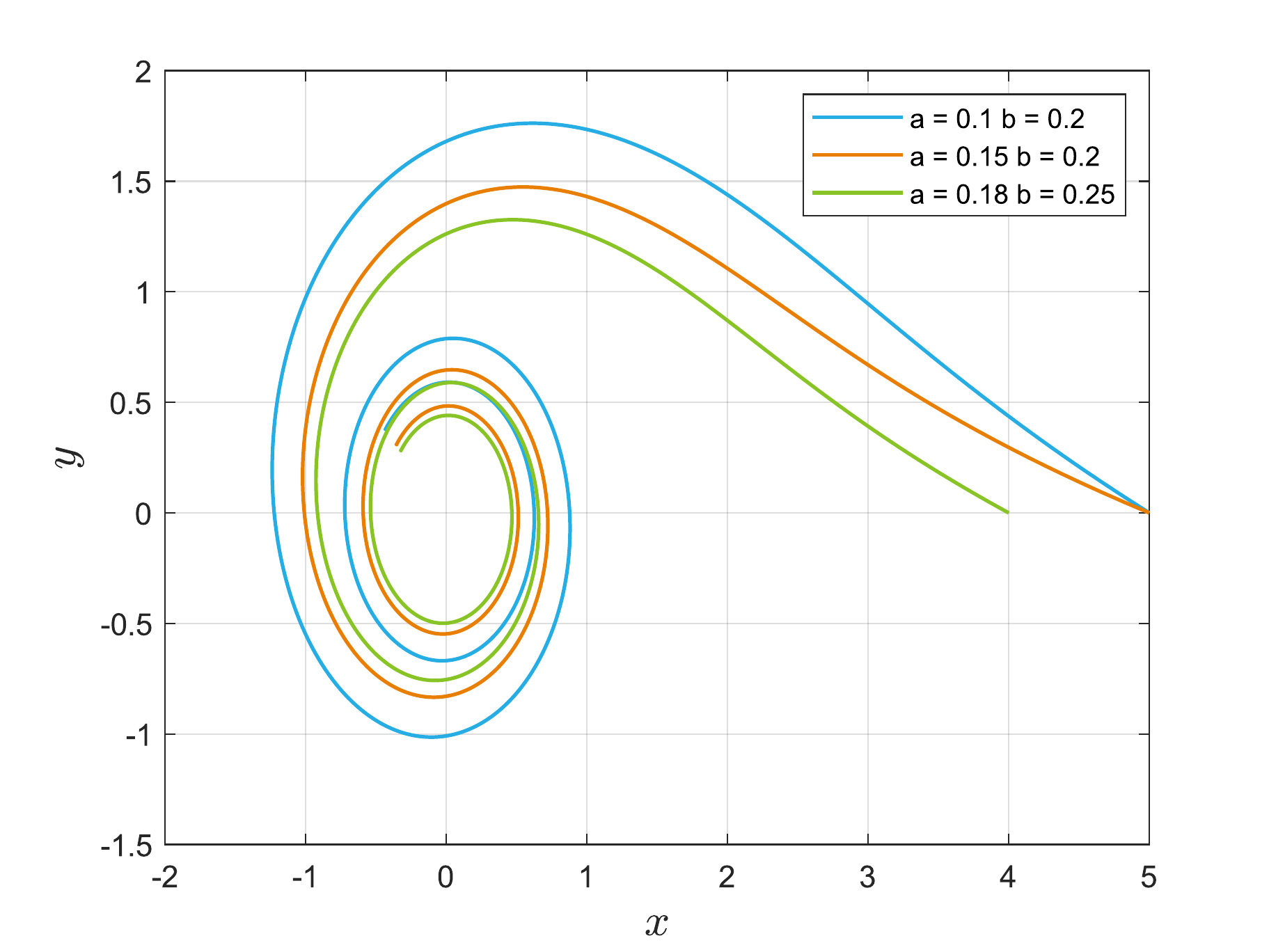}}
\subfigure[Radius of curvature]{\includegraphics[width=0.32\textwidth]{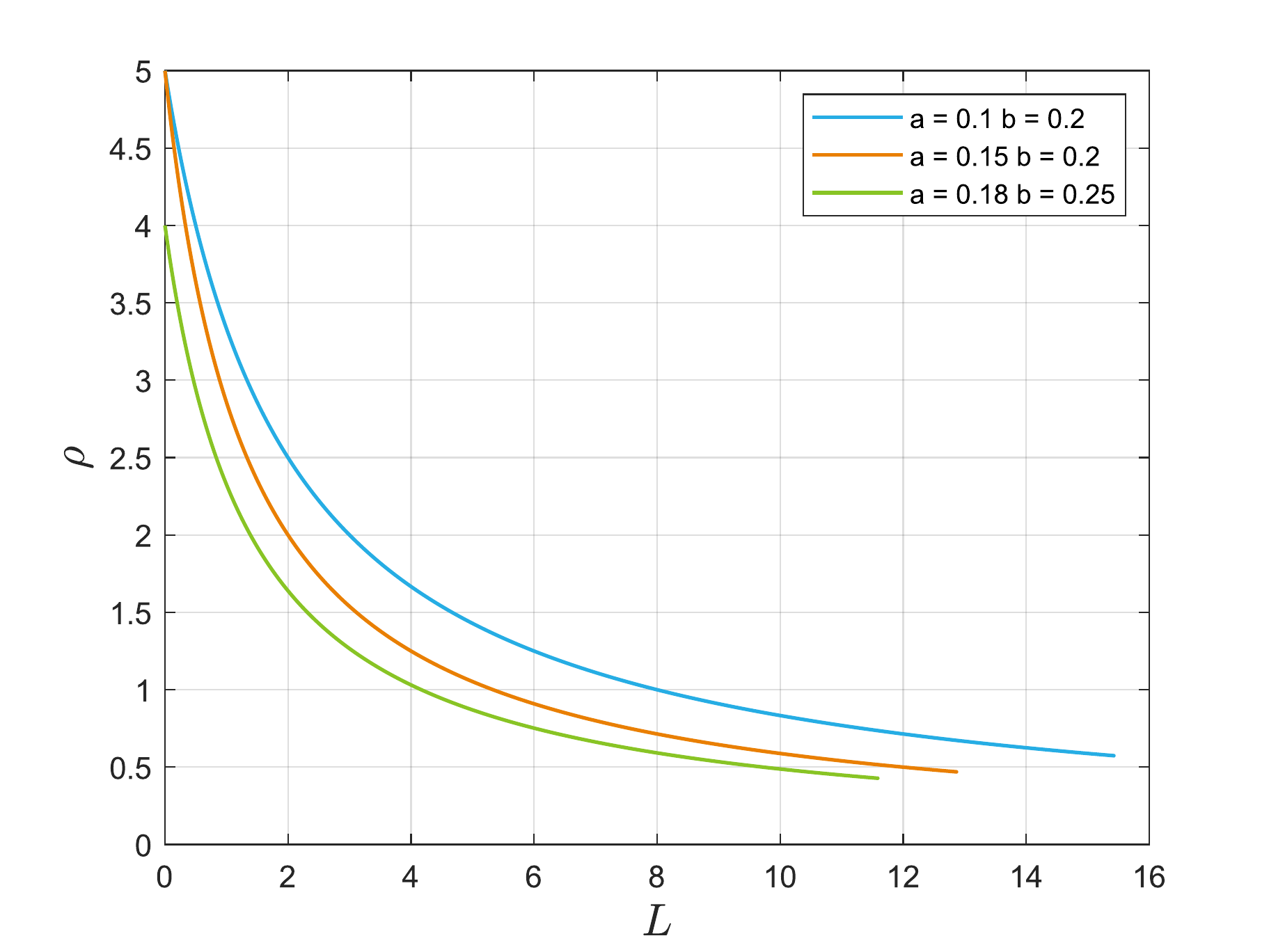}}
\subfigure[Logarithmic curvature]{\includegraphics[width=0.32\textwidth]{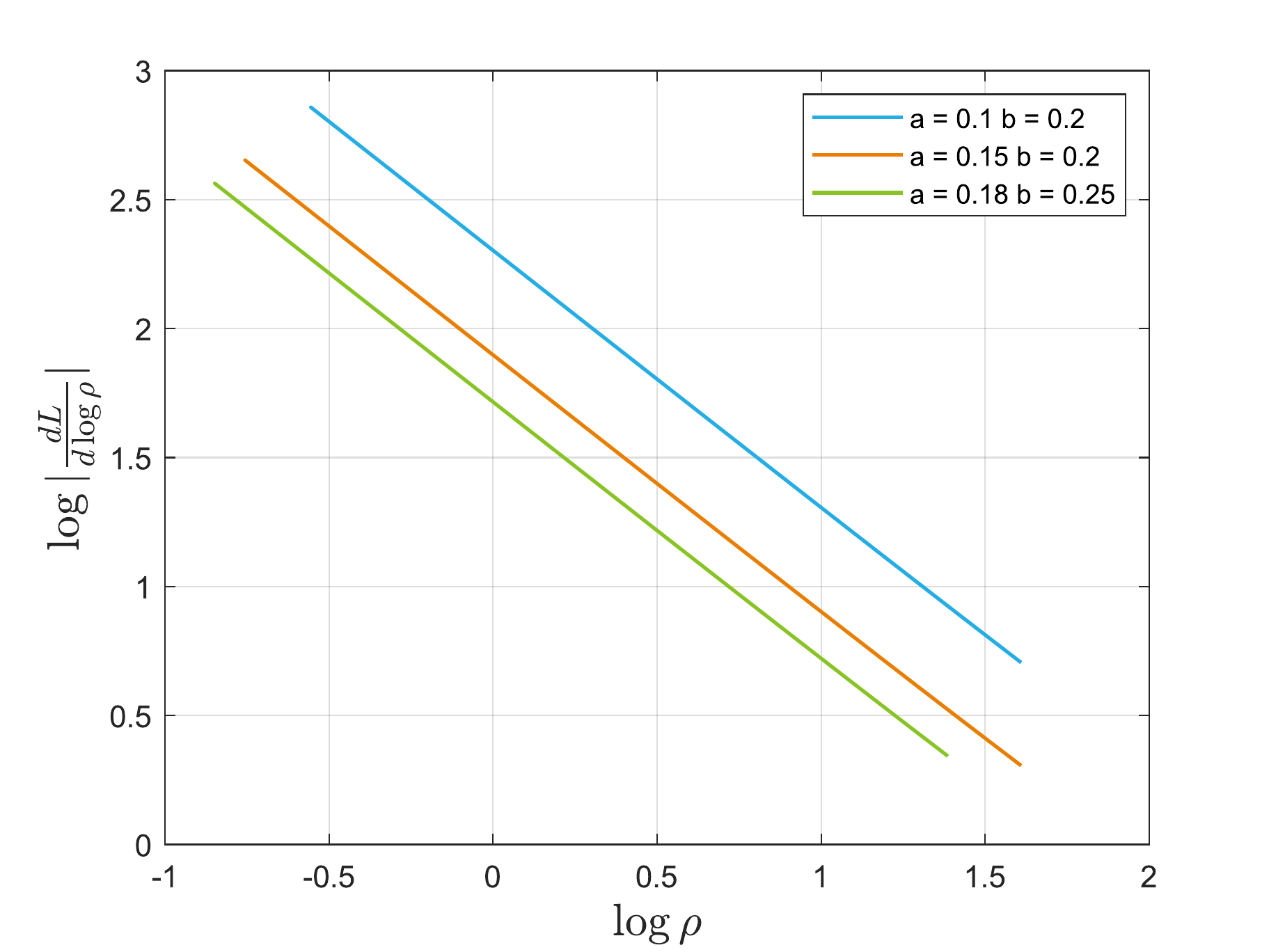}}
\end{center}
\caption{Log aesthetic curves with $n = -1$, $\phi = \pi/2$, $\theta \in [0, 15]$ rad.}
\label{curveB1}
\end{figure*}

\subsubsection{When $\phi$ is constant} Eqn. (\ref{LB}) and Eqn. (\ref{RB}) become

\begin{equation}\label{LB1}
  L = \frac{1}{a} \Bigg ( \Bigg (\frac{a \big (n-1 \big ) \big (\theta - \theo \big ) + nb^{ 1-\frac{1}{n} }}{n} \Bigg )^{\frac{n}{n-1}} - b \Bigg ),
\end{equation}
\begin{equation}\label{RB1}
  R = \big (aL +b \big )^{\frac{1}{n}}  \sin \phi.
\end{equation}

An example of the class of polar aesthetic curves when $n = -1$, $\phi = \pi/2$, $\theta \in [0, 15]$ rad. is provided in Fig. \ref{curveB1}.

\begin{figure*}[t]
\begin{center}
\subfigure[Curve]{\includegraphics[width=0.32\textwidth]{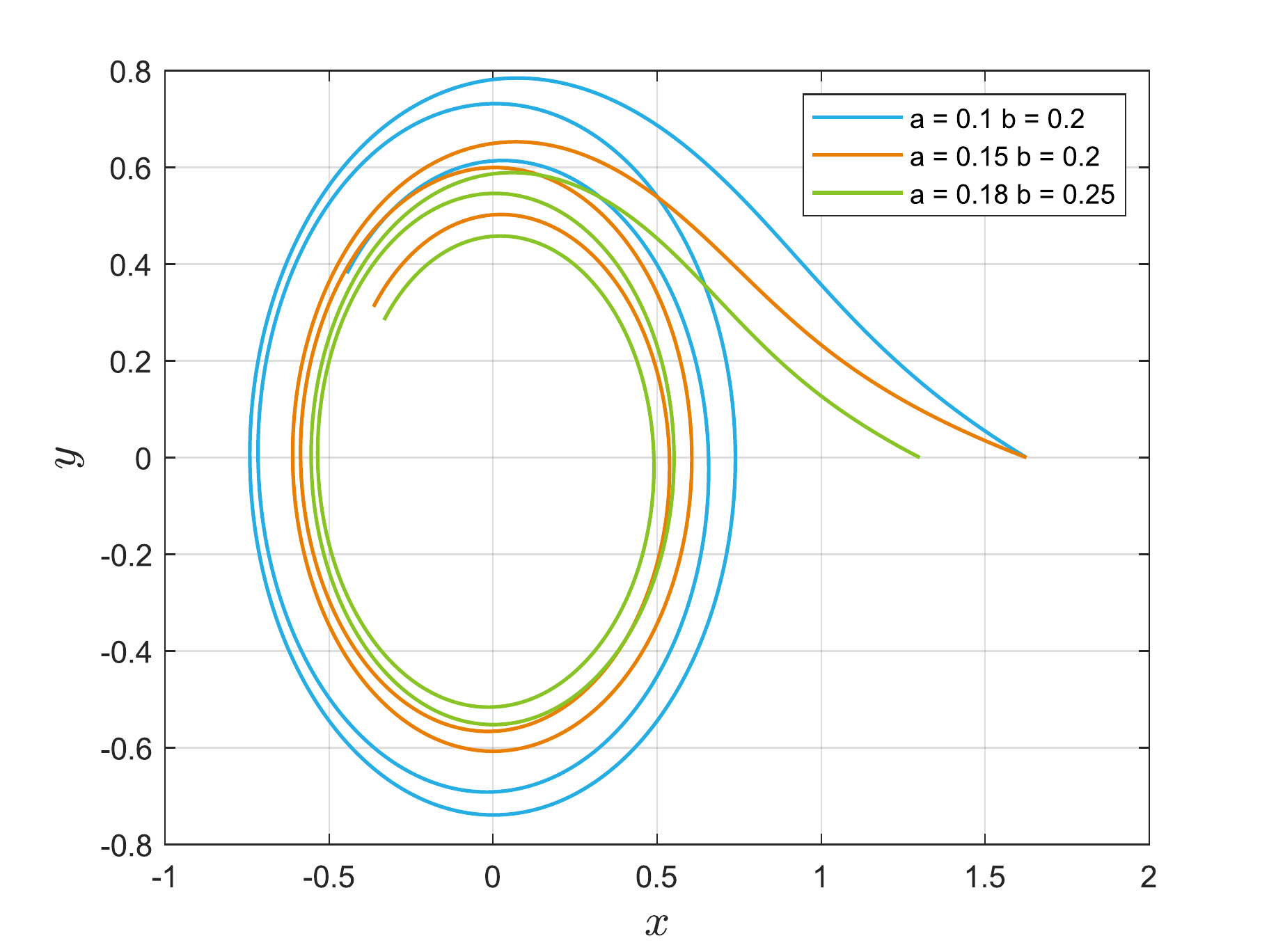}}
\subfigure[Radius of curvature]{\includegraphics[width=0.32\textwidth]{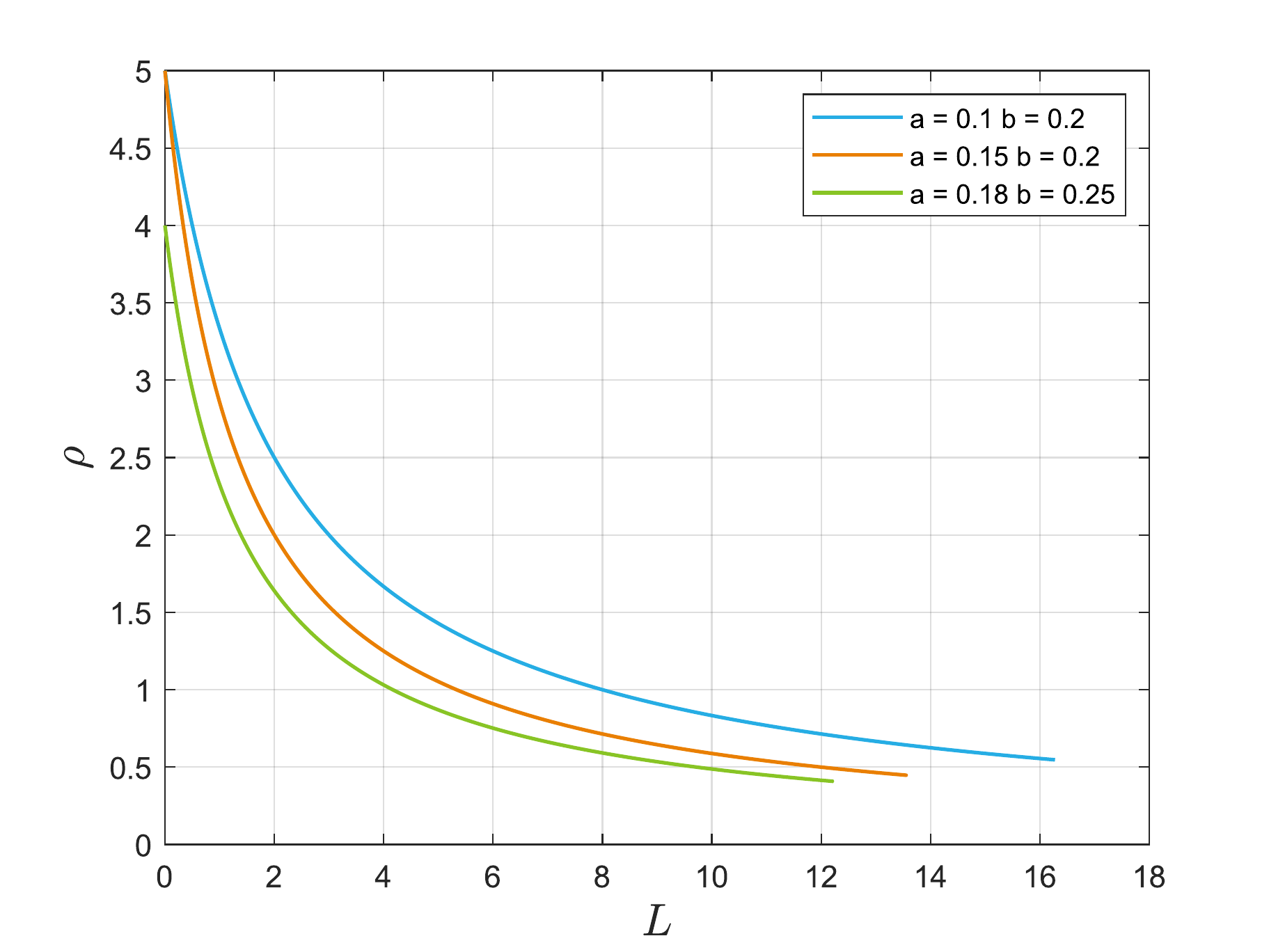}}
\subfigure[Logarithmic curvature]{\includegraphics[width=0.32\textwidth]{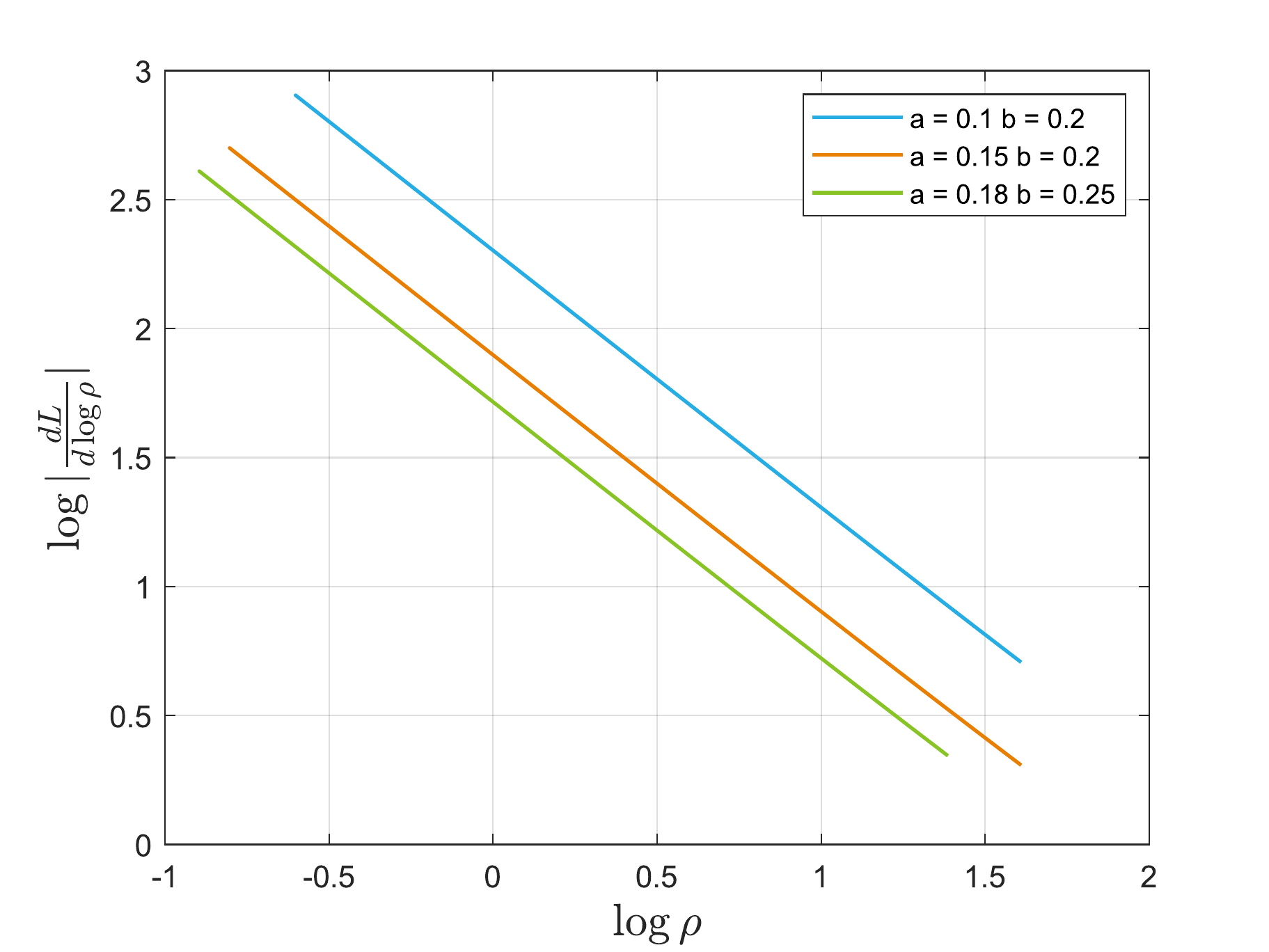}}
\end{center}
\caption{Log aesthetic Curves with $n = -1$, $\phi = 0.1\theta + 0.3$, $\theta \in [0, 15]$ rad.}
\label{curveB2}
\end{figure*}

\subsubsection{When $\phi$ is a linear function of $\theta$}
When $\phi$ is a encoded by $\phi = c_1\theta + c_0 $, Eqn. (\ref{LB}) and Eqn. (\ref{RB}) become

\begin{equation}\label{LA2}
  L = \frac{1}{a} \Bigg ( \Bigg (\frac{a \big (n-1 \big ) \big ( 1+c1\big ) \big (\theta - \theo \big ) + nb^{ 1-\frac{1}{n} }}{n} \Bigg )^{\frac{n}{n-1}} - b \Bigg ),
\end{equation}

\begin{equation}\label{RA2}
  R = \big (aL +b \big )^{\frac{1}{n}}   (1 + c_1 ) \sin (c_1\theta + c_0).
\end{equation}

An example of the class of polar aesthetic curves when $n = -1$, $\phi = 0.1\theta + 0.3$, $\theta \in [0, 15]$ rad. is provided in Fig. \ref{curveB2}.

\begin{figure*}[t]
\begin{center}
\subfigure[Curve]{\includegraphics[width=0.32\textwidth]{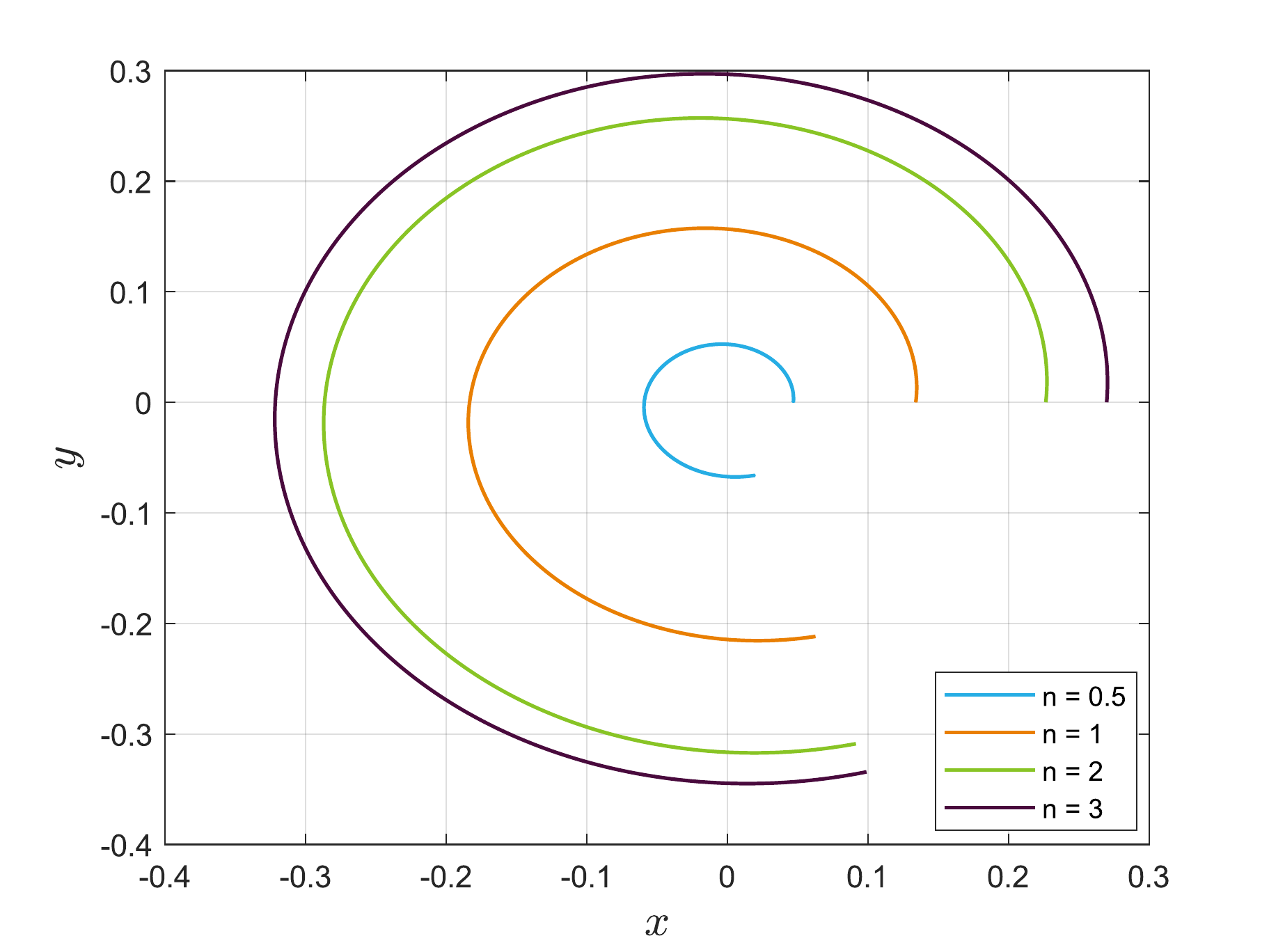}}
\subfigure[Radius of curvature]{\includegraphics[width=0.32\textwidth]{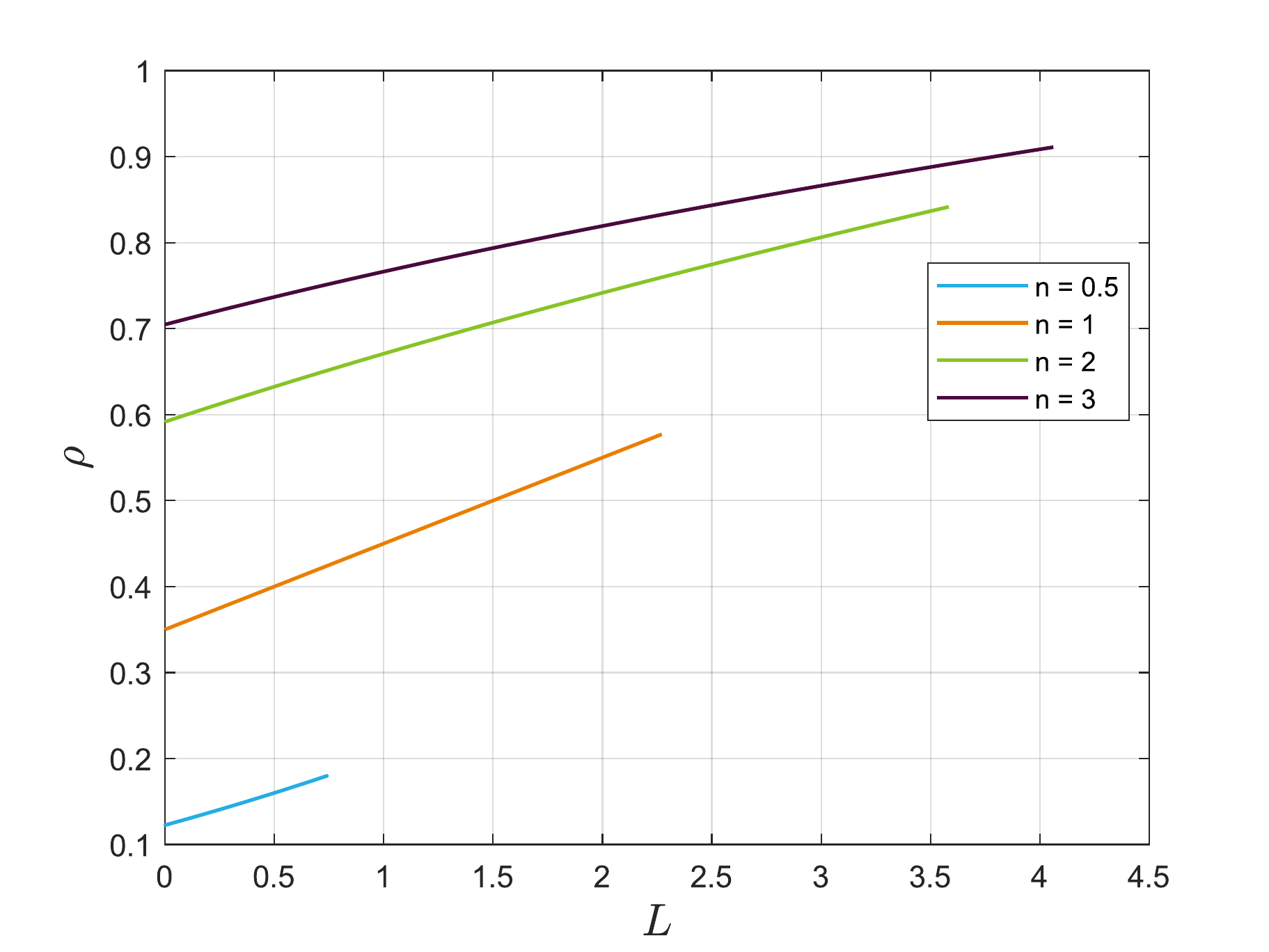}}
\subfigure[Logarithmic curvature]{\includegraphics[width=0.32\textwidth]{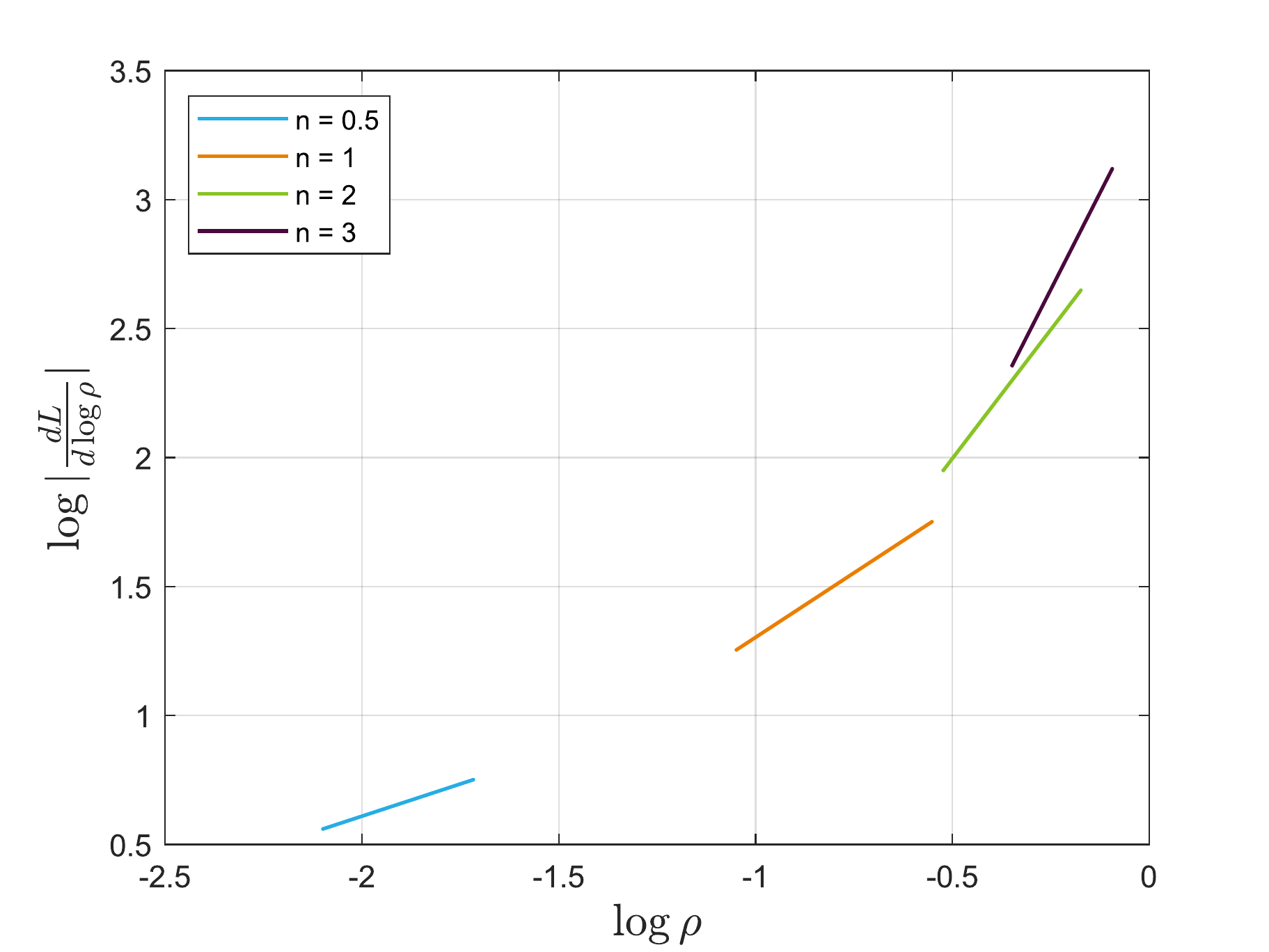}}
\end{center}
\caption{Effect on the positive slope $n$ on the characterization of polar aesthetic Curves when $\phi = \pi/8$, $\theta \in [0, 5]$ rad.}
\label{curveBn1p}
\end{figure*}

\begin{figure*}[t]
\begin{center}
\subfigure[Curve]{\includegraphics[width=0.32\textwidth]{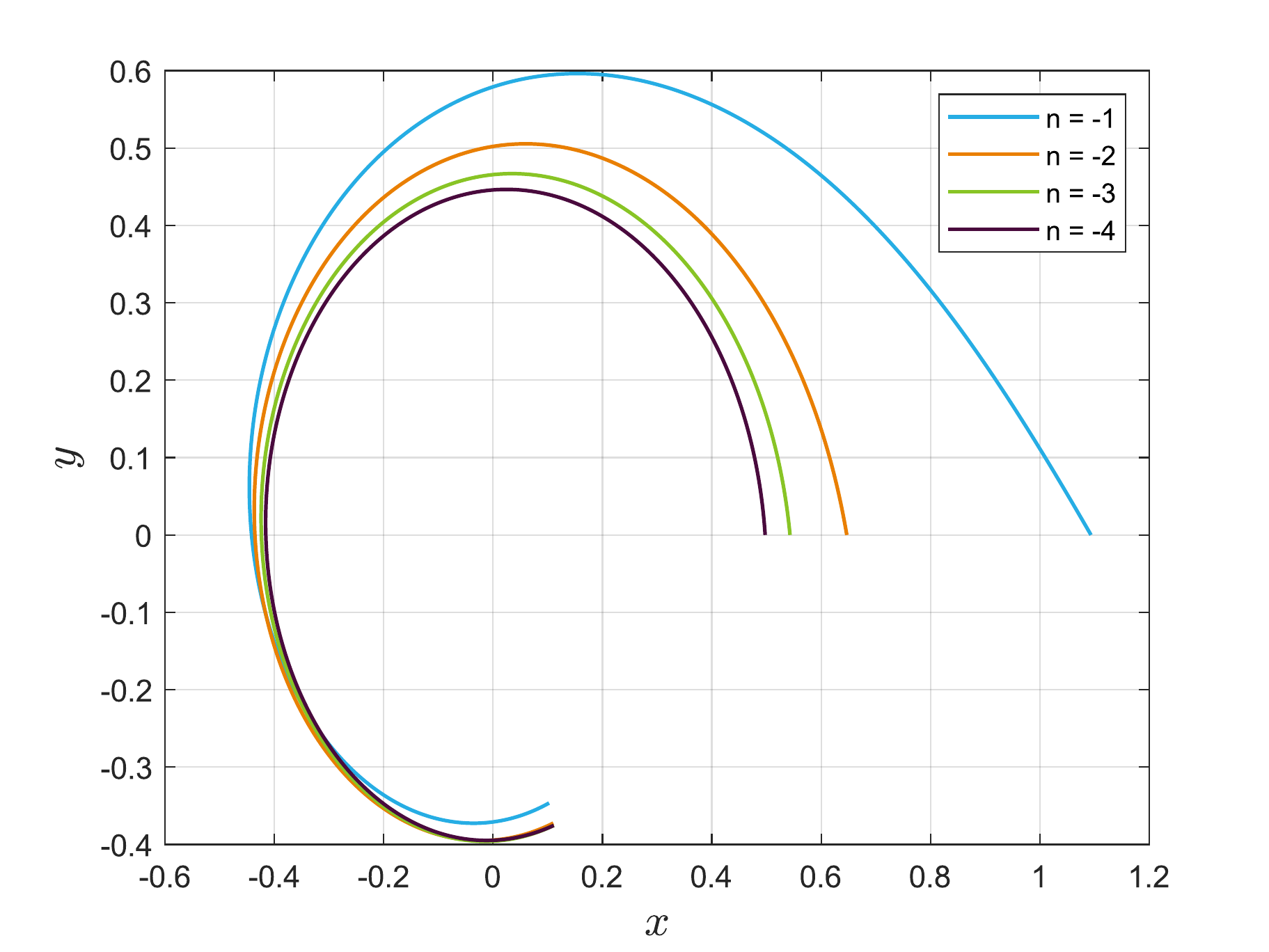}}
\subfigure[Radius of curvature]{\includegraphics[width=0.32\textwidth]{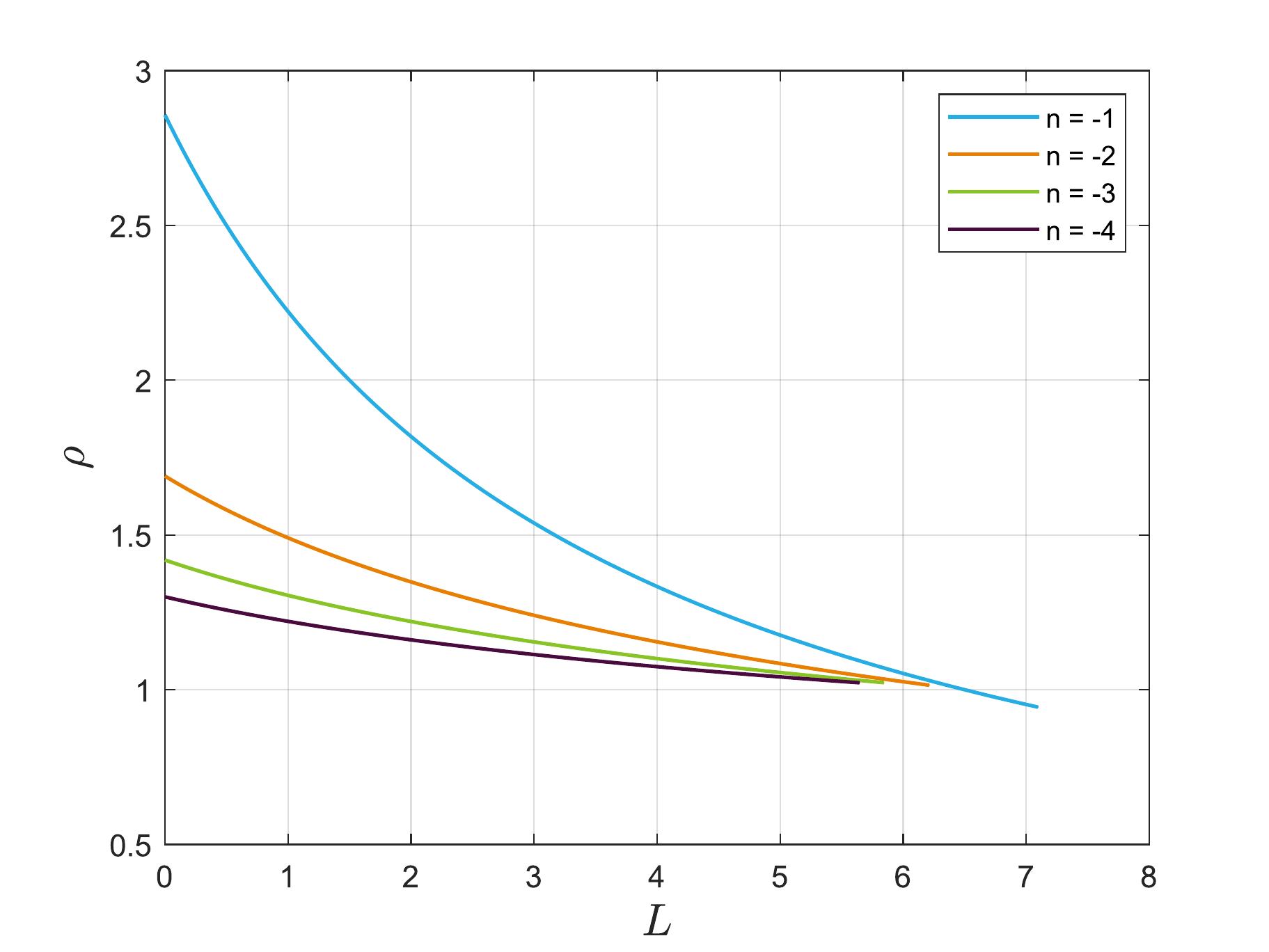}}
\subfigure[Logarithmic curvature]{\includegraphics[width=0.32\textwidth]{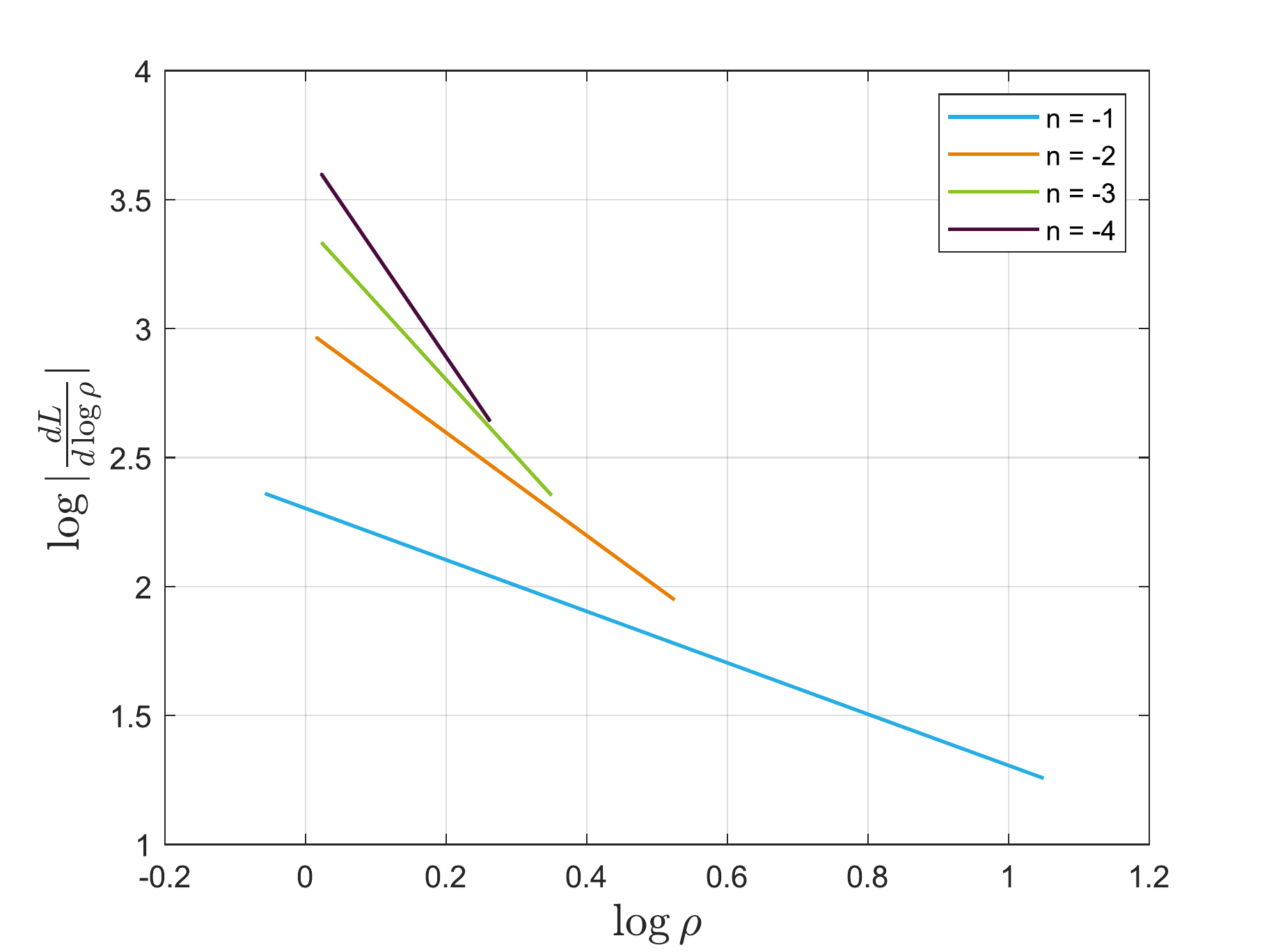}}
\end{center}
\caption{Effect on the negative slope $n$ on the characterization of polar aesthetic Curves when $\phi = \pi/8$, $\theta \in [0, 5]$ rad.}
\label{curveBn1}
\end{figure*}

\begin{figure*}[t]
\begin{center}
\subfigure[Curve]{\includegraphics[width=0.32\textwidth]{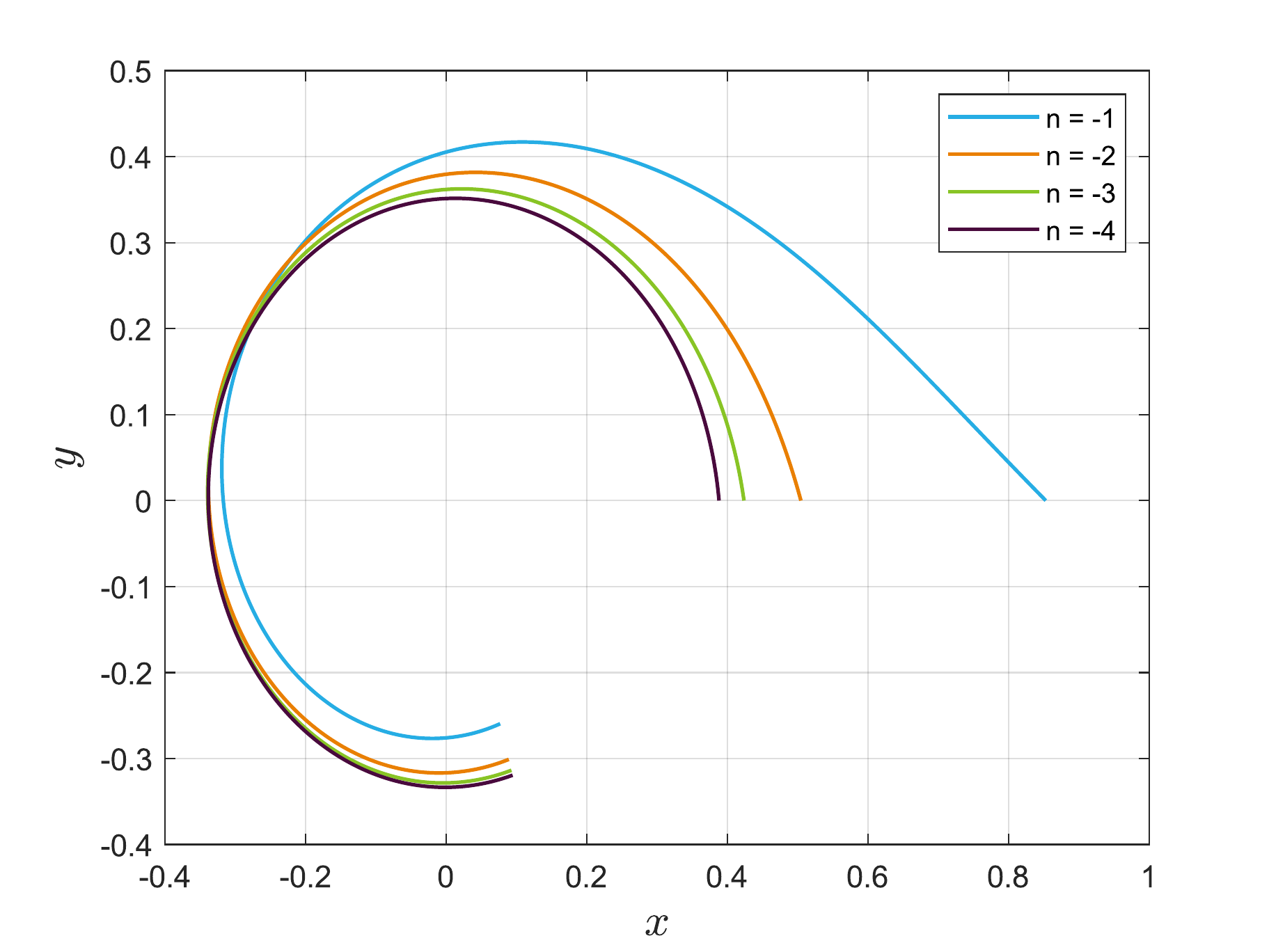}}
\subfigure[Radius of curvature]{\includegraphics[width=0.32\textwidth]{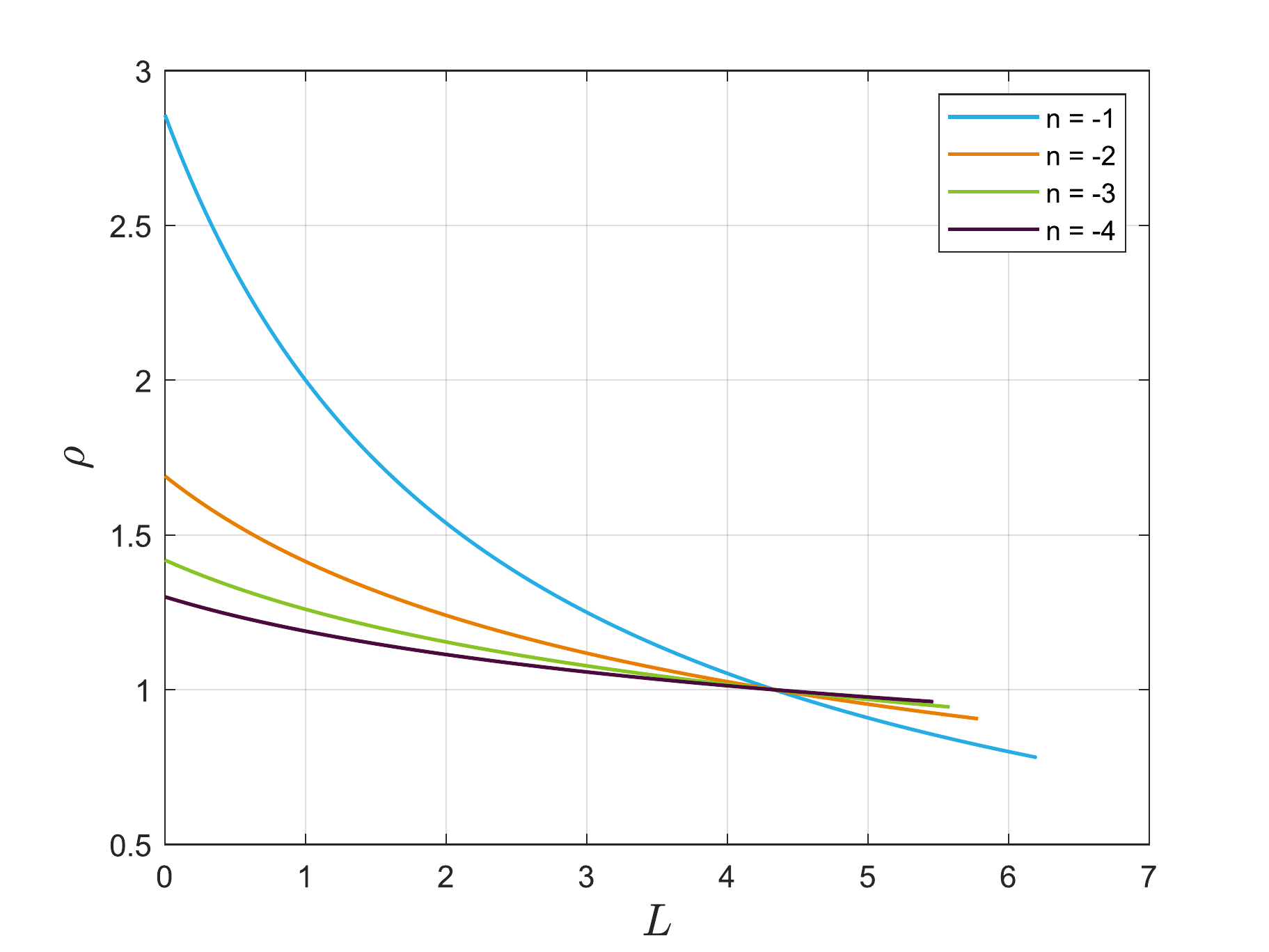}}
\subfigure[Logarithmic curvature]{\includegraphics[width=0.32\textwidth]{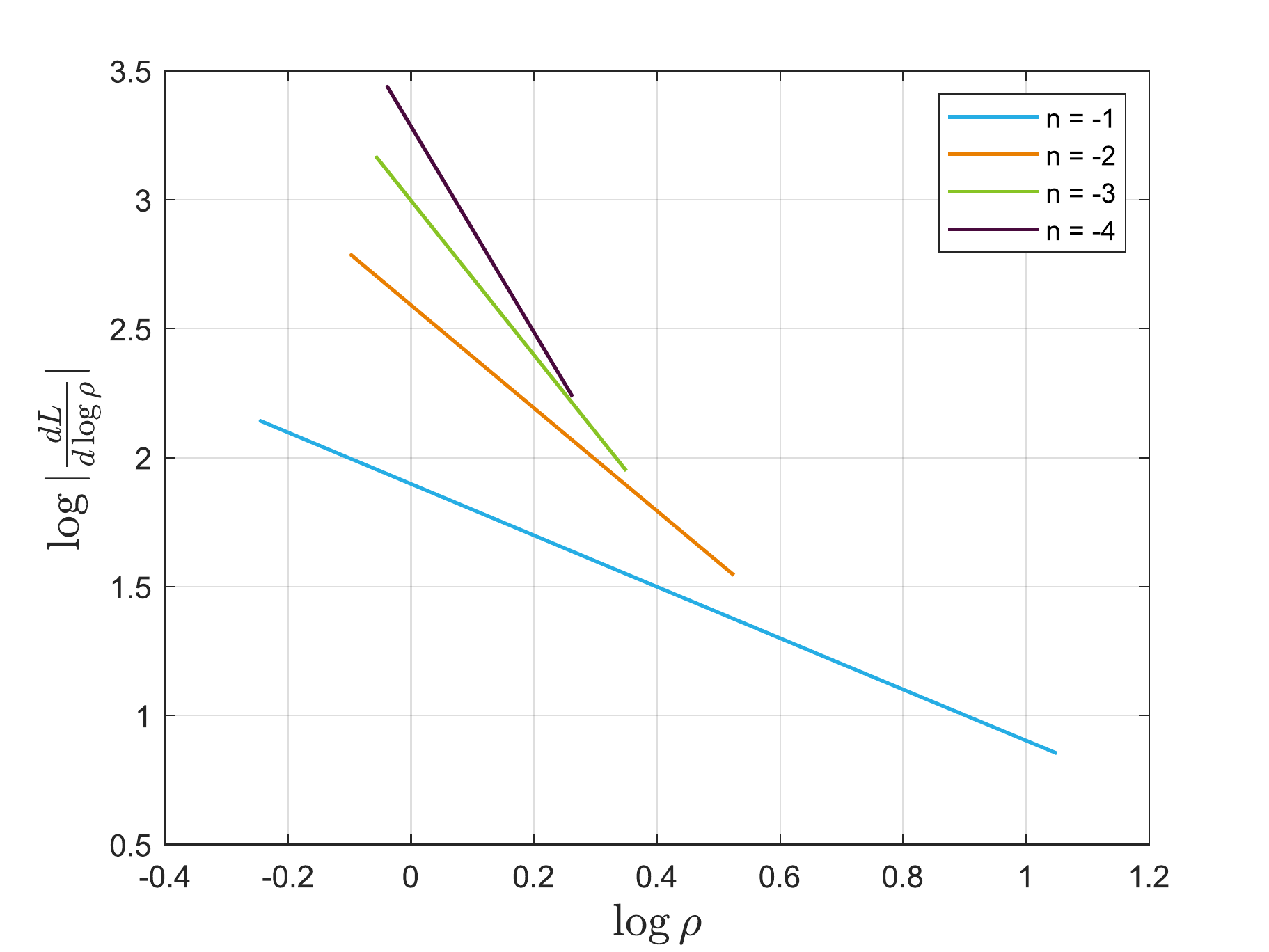}}
\end{center}
\caption{Effect on the negative slope $n$ on the characterization of polar aesthetic Curves when $\phi = 0.01\theta + 0.3$, $\theta \in [0, 5]$ rad.}
\label{curveBn2}
\end{figure*}

\begin{figure*}[h]
\begin{center}
\subfigure[Curve]{\includegraphics[width=0.32\textwidth]{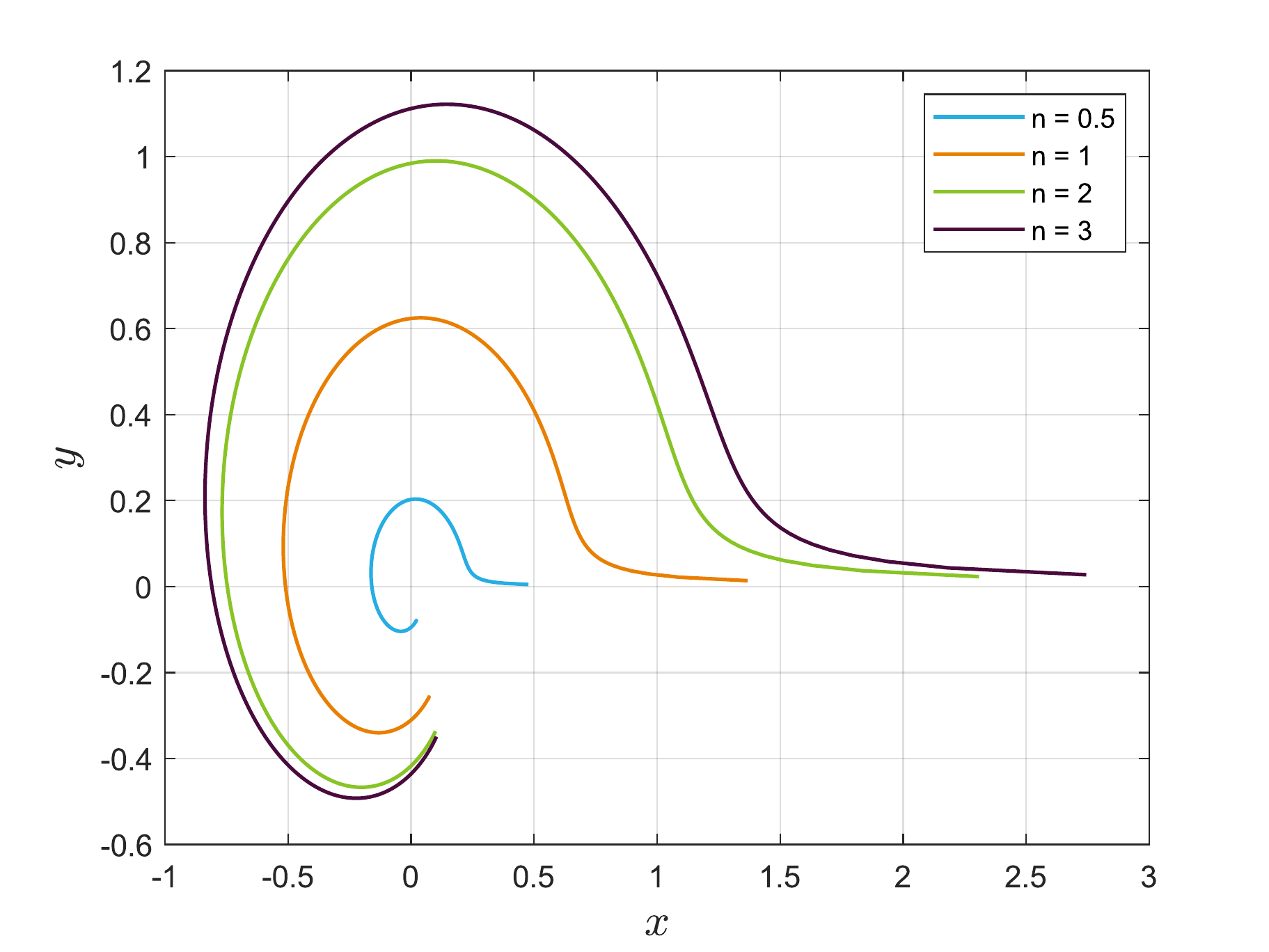}}
\subfigure[Radius of curvature]{\includegraphics[width=0.32\textwidth]{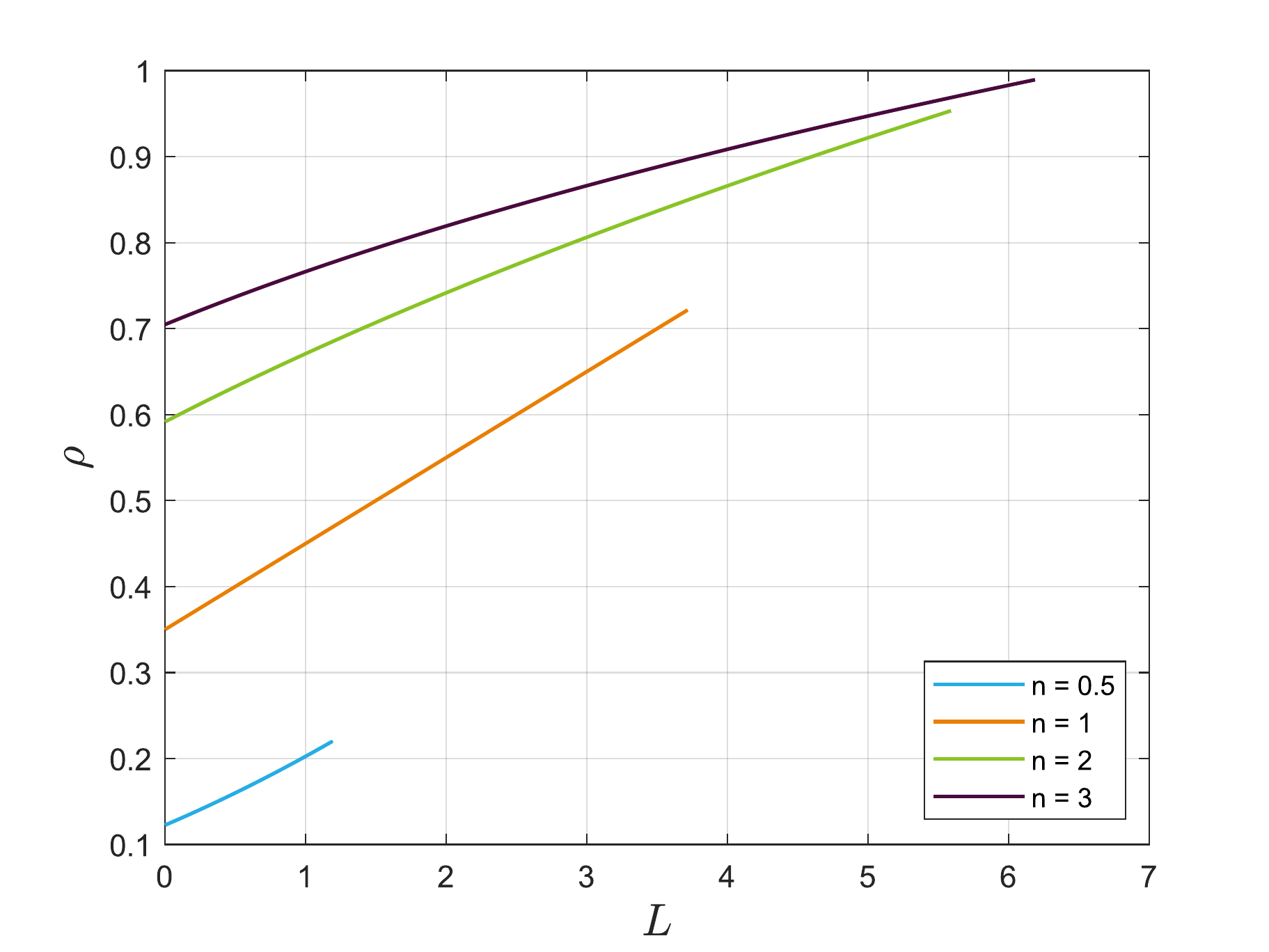}}
\subfigure[Logarithmic curvature]{\includegraphics[width=0.32\textwidth]{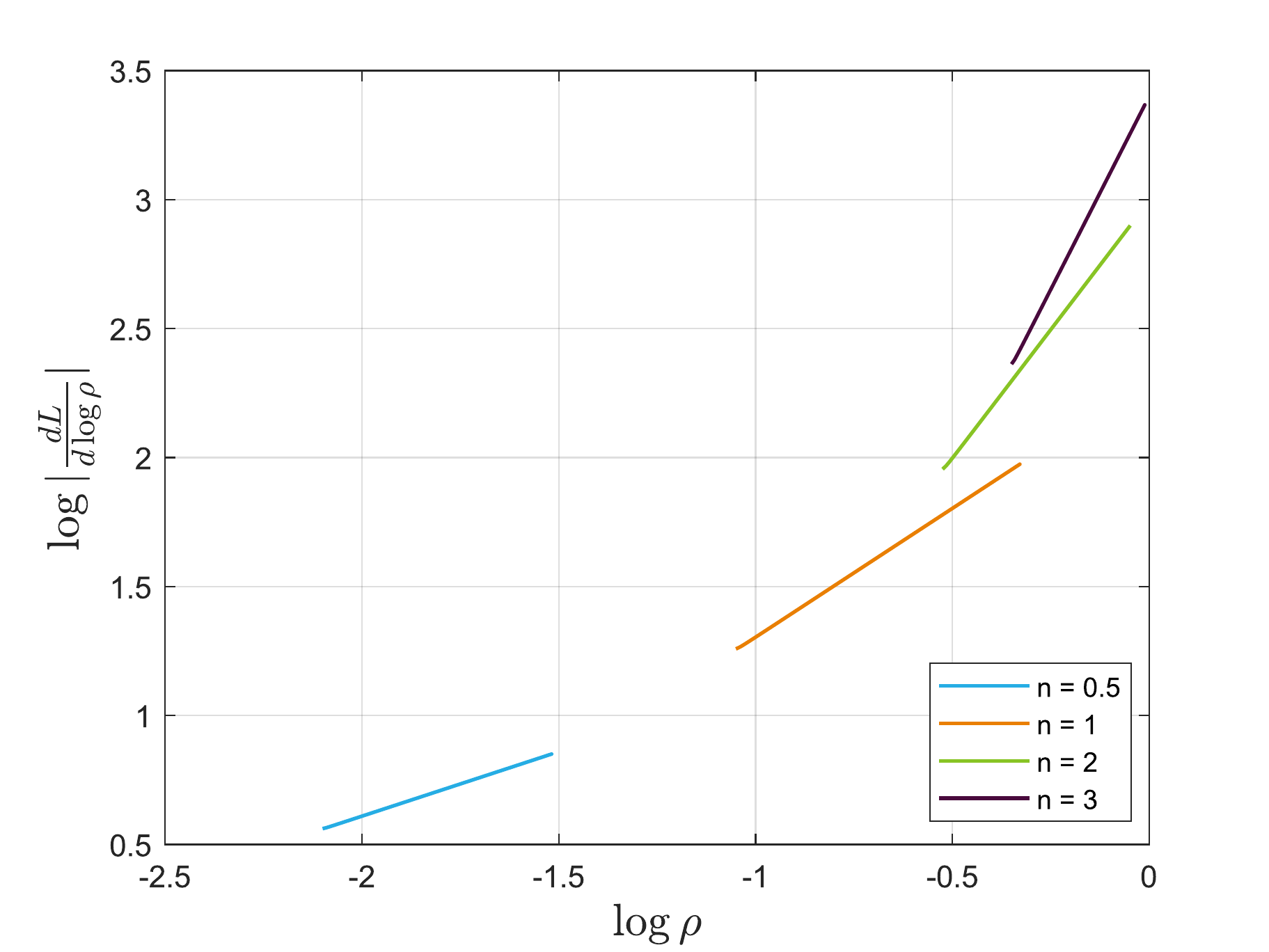}}
\end{center}
\caption{Log aesthetic curves when $\phi = \sqrt{\theta} + 0.6$, $\theta \in [0, 5]$ rad.}
\label{curveBn1pf}
\end{figure*}

By using Eqn. (\ref{LB}) and Eqn. (\ref{RB}), it becomes possible to model distinct polar aesthetic curves for which the logarithmic curvature graph is always linear. For instance, the characterization of polar aesthetic Curves when $\phi = \pi/8$, $\theta \in [0, 5]$ rad. for both positive and negative slopes is presented in Fig. \ref{curveBn1p}, Fig. \ref{curveBn1} and Fig. \ref{curveBn2}. The reader may note that slopes induce in the inward and outward characteristic of the polar aesthetic curves. Naturally, by using Eqn. (\ref{LB}) and Eqn. (\ref{RB}), it also becomes possible to model polar aesthetic curves for which the angle $\phi$ is an arbitrary function, such as the class of curves when $\phi = \sqrt{\theta} + 0.6$ and $\theta \in [0, 5]$ rad. as shown by Fig. \ref{curveBn1pf}.

\section{CONCLUSIONS}

In this paper, we have proposed the closed-form characterizations of polar log-aesthetic curves for given curvature profiles and dynamics of polar angles. We also have presented numerical examples portraying the feasibility to construct polar aesthetic curves whose logarithmic curvature graph is expressed by a controllable straight line. Our approach facilitates the seamless characterization of aesthetic curves in the polar coordinate system, whose applicability is straightforward to model shapes with user-defined curvature profiles.

In future work, we aim at studying the formulations for space curves and their real-world applications. Also, we aim at studying the formulation of compounded aesthetic curves in industrial design, path planning, and navigation systems, e.g., to assemble complex silhouettes and match a specific curvature profile in polar coordinates\cite{smc18,embc18,mohamed19,parquesmooth20, parque20,parque16,higashi14a,higashi14b,higashi15}. To the best of our knowledge, our closed-form characterizations are the first presented in the literature, whose use is potential to design aesthetic curves in CAD/CAE and planning problems.

\bibliographystyle{asmems4}


%
% ...


\bibliography{mybib}

\end{document}